\documentclass[useAMS,usenatbib]{mn2e_astroph}

\usepackage{graphicx}

\def\nct#1{\nocite{#1}}

\newcommand\sss{\scriptscriptstyle}

\def\mix{\psi_{in}}

\def\npsi{N_{\psi,in}}

\def\nlag{N_{\Delta\phi}}

\def\sipsi{\sigma_{\psi,in}}
\def\silag{\sigma_{\Delta\phi}}
\def\psin{\psi_{in}}
\def\lpk{\Delta\phi_{pk}}
\def\ppk{\psi_{pk}}
\def\lag{\Delta\phi}

\title[Coherent radio pulsar polarization]
{Radio pulsar polarization as a coherent sum of orthogonal proper mode
waves}

\author[J.~Dyks]
{J.~Dyks
\\
Nicolaus Copernicus Astronomical Center, Rabia\'nska 8, 87-100, Toru\'n,
Poland\\
}
\begin{document}

\date{Accepted .... Received ...; in original form 2018 Dec 27}


\maketitle

\label{firstpage}

\begin{abstract}
Radio pulsar polarization exhibits a number of complex phenomena that are classified
into the realm of `beyond the rotating vector model' (RVM). It is shown that these
effects can be understood in geometrical terms, as a result of
coherent and quasi-coherent addition of elliptically polarized natural
mode waves. 
The coherent summation implies that the observed tracks of polarization angle (PA)  do not always 
correspond to the natural propagation mode (NPM) waves. Instead, they are statistical average
of coherent sum of the NPM waves, and can be observed at any (and
frequency-dependent) distance from the natural modes. 
Therefore, the observed tracks of PA can wander arbitrarily far from the RVM, 
and may be non-orthogonal.
For equal amplitudes of the NPM waves 
two pairs of orthogonal polarization modes (OPMs), displaced by $45^\circ$, can 
be observed, depending on the width of lag distribution. 
Observed pulsar polarization mainly results from two
independent effects: the change of mode amplitude ratio and the change of phase
lag. In the core region both effects are superposed on each other, which can
produce so complex behaviour as observed in the cores of 
PSR B1933$+$16, B1237$+$25 and J0437$-$4715. Change of the phase lag with
frequency $\nu$ is mostly responsible for the observed strong evolution of these features
with $\nu$. The coherent addition of orthogonal natural waves is a useful
interpretive tool for the observed radio pulsar polarization.
\end{abstract}

\begin{keywords}
pulsars: general -- pulsars: individual: PSR J0437$-$4715 --
pulsars: individual: PSR B1237$+$25 --
pulsars: individual: PSR B1919$+$21 --
pulsars: individual: PSR B1933$+$16 --
radiation mechanisms: non-thermal.
\end{keywords}

\def\lap{\hbox{\hspace{4.3mm}}
         \raise1.5pt \vbox{\moveleft9pt\hbox{$<$}}
         \lower1.5pt \vbox{\moveleft9pt\hbox{$\sim$ }}
         \hbox{\hskip 0.02mm}}

\def\rwobs{R_W}
\def\rwcon{R_W}
\def\rwstr{R_W}
\def\winobs{W_{\rm in}}
\def\woutobs{W_{\rm out}}
\def\phm{\phi_m}
\def\phmi{\phi_{m, i}}
\def\thm{\theta_m}
\def\dres{\Delta\phi_{\rm res}}
\def\win{W_{\rm in}}
\def\wout{W_{\rm out}}
\def\rin{\rho_{\rm in}}
\def\rout{\rho_{\rm out}}
\def\phin{\phi_{\rm in}}
\def\phout{\phi_{\rm out}}
\def\xin{x_{\rm in}}
\def\xout{x_{\rm out}}

\def\thmin{\theta_{\rm min}^{\thinspace m}}
\def\thmax{\theta_{\rm max}^{\thinspace m}}

\section{Introduction}
\label{intro}

Radio pulsars exhibit a wealth of polarization phenomena that have been
studied for half a century.
However, both the regular polarization properties as well as peculiar effects
escape thorough understanding. The regular behaviour includes the appearance of two orthogonal polarization modes
(OPMs) and transitions (jumps) between these OPMs at several longitudes
in a pulse profile. Peculiar effects are numerous and involve strong deformations of
polarization angle (PA) curve, especially at the central (core) profile components
(Smith et al.~2013, hereafter SRM13; Mitra et al.~2015, hereafter MAR15) 
\nct{mar2015, srm13}
as well as `half orthogonal' PA jumps (Everett \& Weisberg 2001; MAR15).
\nct{ew2001, mar2015}
The research on the subject includes the 
analysis of the natural propagation
wave modes in magnetised plasma (Melrose 1979; Lyubarskii \& Petrova 1999;
Rafat et al.~2018), 
\nct{m79, lp98, rmm2018} 
curvature radiation properties (Gangadhara 2010), \nct{g10}
numerical polarized ray tracing
(Wang et al.~2010), \nct{wlh10}
coherent (Edwards \& Stappers 2004) \nct{es04} and noncoherent deconvolution into separate modes
(Melrose et al.~2006; McKinnon 2003), 
\nct{mmk2006, mck2003} instrumental noise effects
(McKinnon \& Stinebring 2000) as well as interstellar propagation effects
(Karastergiou 2009; McKinnon \& Stinebring 1998). 
\nct{ms98, ms00, k2009}
This is accompanied by a steady increase in the
available polarization data of ever increasing quality (eg.~recently Rankin
et al.~2017; Brinkman et al.~2018). 
\nct{bfrs2018, rah2017}

In this paper I develop the polarization model based on
coherent addition of waves in two orthogonal propagation modes (Dyks 2017,
hereafter D17). The extended model offers a more general nature of 
the observed PA tracks and solves several interpretive obstacles that have appeared in D17. 
\nct{d2017}

In Sect.~\ref{inspiration} I describe observations and modelling hints
that inspired this study. These suggest the importance of equal modal amplitude
in pulsar signal,
so in Sect.~\ref{model} I describe a special-case model based on coherent addition of
linearly polarized waves of equal amplitude. The model is used to interpret
observations in Sect.~\ref{interp}, which is a good opportunity to
present the model properties. Since the equal modal amplitudes may be driven by a
circularly polarized signal, in Sect.~\ref{digression} 
the model based on the
circular feeding is extended into a `birefringent filter pair' model which is
applied to the issue of why the OPMs are so often observed
nearly equal. Section \ref{twofold} describes the double, ie.~convolved or mixed nature 
of the polarization observed in the core profile region such as demonstrated by the
case of PSR J0437$-$4715. The equal amplitudes and linear polarization of the natural mode waves 
cause some interpretive problems 
(described in Sect.~\ref{towards}), therefore, the ellipticity and different
amplitudes of the modal
waves are taken into account in a more general model described in Sect.~\ref{general}. A
glimpse of the properties of the model's parameter space is
 given in Sect.~\ref{lagpa}. Interpretive capabilities of the model are
presented in Sect.~\ref{paloop} where the PA loop of PSR B1933$+$16 is modelled at two
frequencies.

\section{Inspiring observations and modelling hints}
\label{inspiration}

\subsection{Observations}
\label{obs}

\begin{figure}
\includegraphics[width=0.48\textwidth]{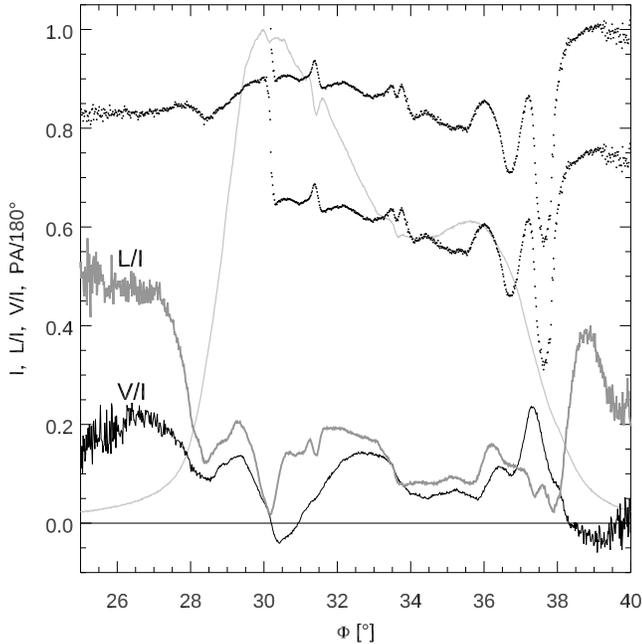}
\caption{Polarisation characteristics of PSR B1919$+$21 at 352 MHz, after MAR15.
The dots present the PA in units of $180^\circ$, plotted at an arbitrary
absolute value. The PA is plotted twice at a distance of $45^\circ$
(equivalent to $0.25$). Light grey intensity profile is shown in the
background for reference. The PA jump near $\Phi=30^\circ$ is nearly 
perfectly equal to $45^\circ$. The circular polarization fraction (bottom black
solid line) passes through zero at the jump and $L/I$ (grey) is minimal.
}
\label{pul45}
\end{figure}

\begin{figure}
\includegraphics[width=0.48\textwidth]{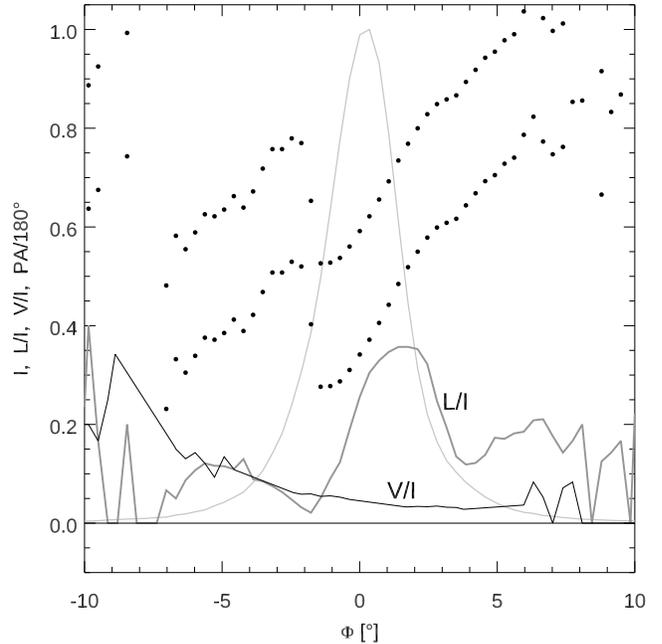}
\caption{Polarization characteristics of PSR B0823$+$26 at $1.4$ GHz, after
Everett \& Weisberg (2001). The same convention as in Fig.~\ref{pul45} is
used, with the PA (dots) plotted twice at a distance of $45^\circ$. The
$45^\circ$ jump at $\Phi\approx2^\circ$ coincides with a minimum in $L/I$, but
the circular polarization fraction (black solid) is not 
affected.
}
\label{ew45}
\end{figure}

Fig.~\ref{pul45} presents the polarized profile of PSR B1919$+$21 as observed by MAR15.
\nct{mar2015}
The profile exhibits a sharp $45^\circ$ PA jump near the maximum flux in the profile. 
The PA is plotted twice at a separation of $45^\circ$ which shows that the
change of PA at the jump is near perfectly equal to $45^\circ$. This may seem
not strange given that the jump coincides with deep minimum in the linear
polarization fraction $L/I$ and with a sign change of the circular polarization
fraction $V/I$. These are trademark features of the equal modal power, and
are frequently observed at the regular OPM jumps. However, after some
wiggling on the trailing side of
the profile, at the pulse longitude $\Phi=37.5^\circ$ the PA makes another
$45^\circ$ downward transition, quickly followed by a more standard $90^\circ$
upward OPM jump at $\Phi=38^\circ$. 

When moved up by $45^\circ$, the displaced central PA segment  (between $\Phi=30^\circ$ and
$37^\circ$) provides roughly rectilinear interpolation
between the PA observed outside of the segment. This suggests that 
the PA stays at the $45^\circ$ distance through most of the pulse window, 
and there must be some geometric reason for this. The $45^\circ$ shift seems to
exist despite a clearly
nonzero level of both $L/I$ and $V/I$. Fig.~18 in MAR15 shows that a
chaotic multitude of different PA values are observed 
within the displaced-PA interval of pulse longitude.  

As can be seen in Fig.~\ref{ew45}, based on Fig.~7 in Everett \& Weisberg (2001),
\nct{ew2001} 
PSR B0823$+$26 also shows a $45^\circ$ jump which is coincident with a
minimum in $L/I$. In this case, however, the
profile of $V/I$ does not seem to be affected by the phenomenon.

In D17 the half-orthogonal PA jump has been interpreted as a sudden narrowing of
a phase delay distribution, with the delay measured between two linearly
polarized waves, supposedly representing the waves of natural propagation
modes. The small delays imply coherent addition of waves, which ensures the
$45^\circ$ PA jump as soon as the waves have equal amplitudes. However, this
rises interesting questions. First, what makes the amplitudes equal, and
second --  having two pairs of orthogonal PA values off at $45^\circ$ -- which pair 
coincides with the PA of the supposedly quasi linear\footnote{Hereafter, the
terms `linear' or `circular', when referring to signals, waves or modes, should
always be understood as `linearly polarized' and `circularly polarized',
respectively.}  
natural polarization modes?
Below I will confirm the idea of the lag-distribution narrowing, however,
the identification of the modes will be shown to depend on whether
equal modal amplitudes can be sustained in pulsar signal.

\begin{figure}
\includegraphics[width=0.48\textwidth]{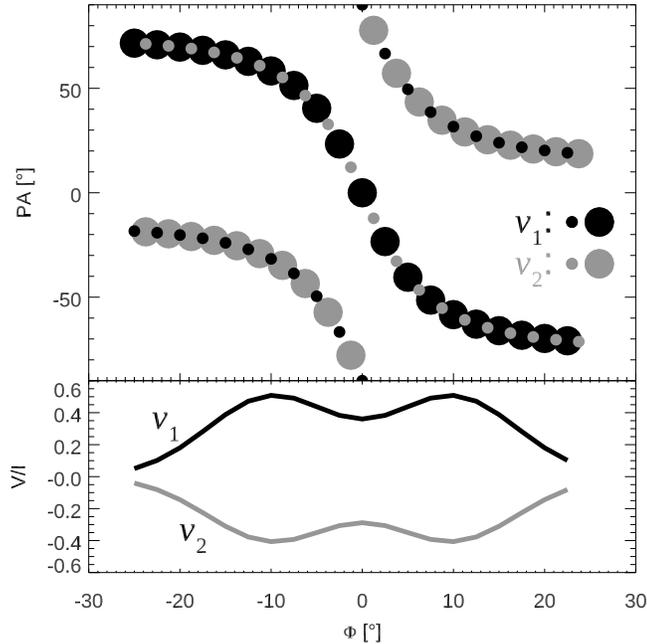}
\caption{Cartoon presentation of the primary mode exchange effect, 
after Figs.~4 and 7 of Young \& Rankin (2012). The black and grey colors refer
to different frequencies $\nu_1$ and $\nu_2$. The size  of points represents
the strength of modal track. With the change of $\nu$ the strongest mode
appears at orthogonal position ($90^\circ$ off) and the observed $V$ changes
sign to opposite (bottom panel).
}
\label{repla}
\end{figure}

Another type of interesting polarization phenomenon is the exchange of
the observed modal power with increasing frequency $\nu$. This is well
illustrated in Figs.~4 and 5 of Young and Rankin (2012) 
\nct{yr12}
where single pulse
PA distributions are shown at two frequencies for PSR B0301$+$19 and
B1133$+$16.\footnote{The authors do not comment on this exchange at all.}
In Fig.~\ref{repla} I show a cartoon representation of this effect.  
The PA distribution in each pulsar reveals two enhanced PA tracks that follow 
a pair of well defined rotating vector model (RVM) curves, 
with each PA track apparently representing a different OPM.   
However, the primary (ie.~brighter) mode track at 327 MHz becomes the
secondary (fainter) track at 1.4 GHz.  
According to the authors, the data were corrected for the interstellar
Faraday rotation, various instrumental effects and dispersion. Moreover,  
the apparent replacement of the modal power is confirmed by 
a probably concurrent
change
 in the sign of $V$. The power of the observed OPMs is then partially separated not only in
pulse longitude and drift phase (Edwards \& Stappers 2003; Rankin \&
Ramachandran 2003; Edwards 2004), 
\nct{es2003, edw2004, rr03}
but also in the spectral domain (Noutsos et al.~2015). \nct{nsk15}
This may seem to be natural, because the modes are generally expected to have
different refraction indices, each with different dependence on $\nu$, which
implies a $\nu$-dependent phase lag between the modal waves.
In the model of D17, however, any changes of the phase lag could not affect the
ratio of modal power. This lack of flexibility makes the $\nu$-related
considerations difficult and calls for the model extension.

\begin{figure}
\includegraphics[width=0.48\textwidth]{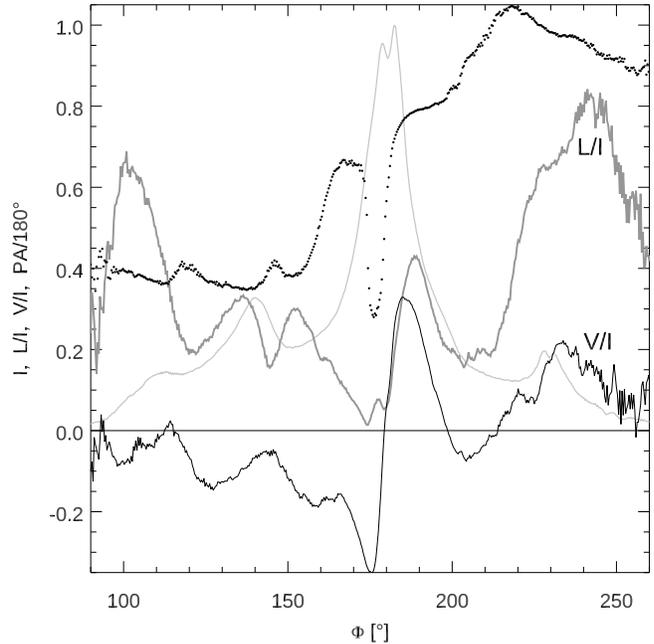}
\caption{Polarization characteristics of J0437$-$4715 at 660 MHz, after
Navarro et al.~(1997). Graphical convention is the same as in
Fig.~\ref{pul45}, except from that the PA (dots) is plotted once. The
profile of $L/I$ has two nearby minima in the profile center ($\Phi
=175^\circ$). The right minimum coincides with the handedness change of $V$. 
The left one occurs at high $V/I$. Both minima correspond to OPM jumps
visible in the PA (the right one has magnitude close to $90^\circ$).
}
\label{j0437}
\end{figure}

Another type of insightful polarization effects are the distortions and bifurcations of
polarization angle tracks, especially those observed within the central
(core) components of pulsars such as PSR B1237$+$25 (SRM13), B1933$+$16
(Mitra et al.~2016, hereafter MRA16), B1857$-$26 (Mitra \& Rankin 2008), and B1839$+$09 (Hankins \& Rankin
2010). 
\nct{srm13, mra2016, mr2008, hr10}
All these phenomena reveal clear signatures of
their coherent origin: they have maxima of $|V|$ coincident with minima of
$L/I$. The loop-like PA distortion of B1933$+$16 
was modelled in section 4.4.1 in D17, whereas the PA track
bifurcation of B1237$+$25 was interpreted in section 4.7 therein. 
Here those interpretations will be modified and will be made consistent with
each other.

\begin{figure*}
\includegraphics[width=0.78\textwidth]{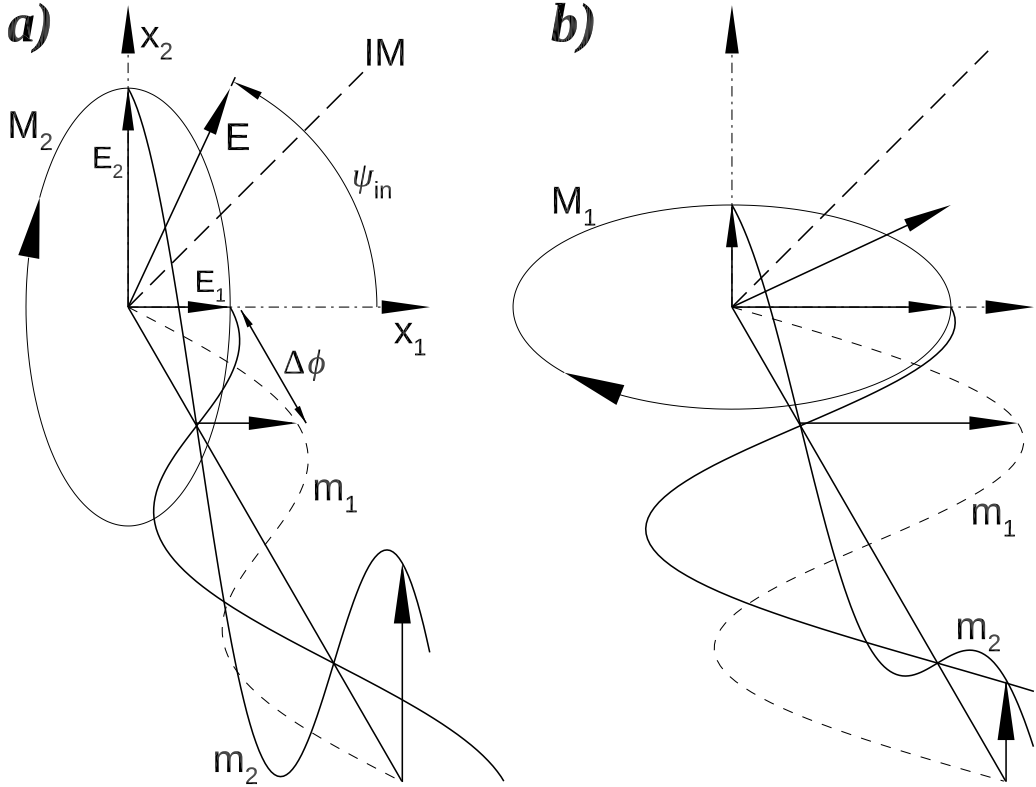}

\caption{The origin of observed coproper OPMs (represented by the ellipses M$_1$ and M$_2$) as the coherent 
sum of phase-lagged proper mode waves m$_1$ and m$_2$. The linearly polarized proper waves
m$_1$ and m$_2$ are fed by the linearly polarized wave $\vec E$, which enters a 
linearly birefringent medium at the mixing angle $\psin$.  The proper
polarization directions of the medium are presented by $\vec x_1$ and $\vec
x_2$, $\lag$ is the phase lag acquired from refraction index difference, and
IM is the intermodal separatrix at $\psin=45^\circ$ (the crossing of which
corresponds to the pseudomodal 
OPM transition). The phase lagged position of the proper wave m$_1$ 
is shown with the dashed line.  For the selected $\psin$, the observed OPMs (M$_1$ and M$_2$)
have the same handedness.
}
\label{prinz}
\end{figure*}

An interesting example of the core PA distortion is provided by the millisecond
pulsar PSR J0437$-$4715 (Navarro et al.~1997, hereafter NMSKB97, Oslowski et
al.~2014). 
\nct{nms97, ovb14} 
As shown in Fig.~\ref{j0437} (after NMSKB97), at 660 MHz the PA curve
steeply dives to the vicinity of orthogonal mode, then immediately retreats
in
another nearly full OPM jump. The retreat is associated with the sign change
of $V$ and a minimum in $L/I$ (which is not quite vanishing).
The first quasi-OPM transition, however, is associated with a high level of
$|V|/L$. Section 4.4 in D17 describes an effect of symmetric twin minima in $L/I$ which
are associated with symmetric profile of $V/I$. Both these
minima have identical look and identical origin. In PSR J0437$-$4715, however,  
the observed minima are dissimilar and have clearly different
origin.  
Moreover, when viewed at different frequencies (NMSKB97) the minima 
seem to move in longitude at a different rate. They seem to pass across each other which
is apparently related to $\nu$-dependent amplitudes of the negative and positive $V$, 
and is accompanied by strong changes of PA distortions. 
Overall, the behaviour of polarization in PSR J0437$-$4715 looks as a clear
manifestation of two independent processes that overlap in pulse longitude. 

Another strange polarization effect can be seen on the trailing side of the
core component in J0437$-$4715 (Fig.~\ref{j0437},
$\Phi\in(200^\circ,220^\circ)$). The PA there seems to be freely
wandering with no obedience to any RVM-like curve. Off-RVM PA values must
also be involved in a phenomenon of non-orthogonal PA tracks, that is
often observed in many pulsars (eg.~B1944$+$17 and B2016$+$28, both at $1.2$
GHz in Fig.~15 of MAR15). Apparently, any successful pulsar polarization
model must be capable of easily detaching from RVM.

\begin{figure*}
\includegraphics[width=0.78\textwidth]{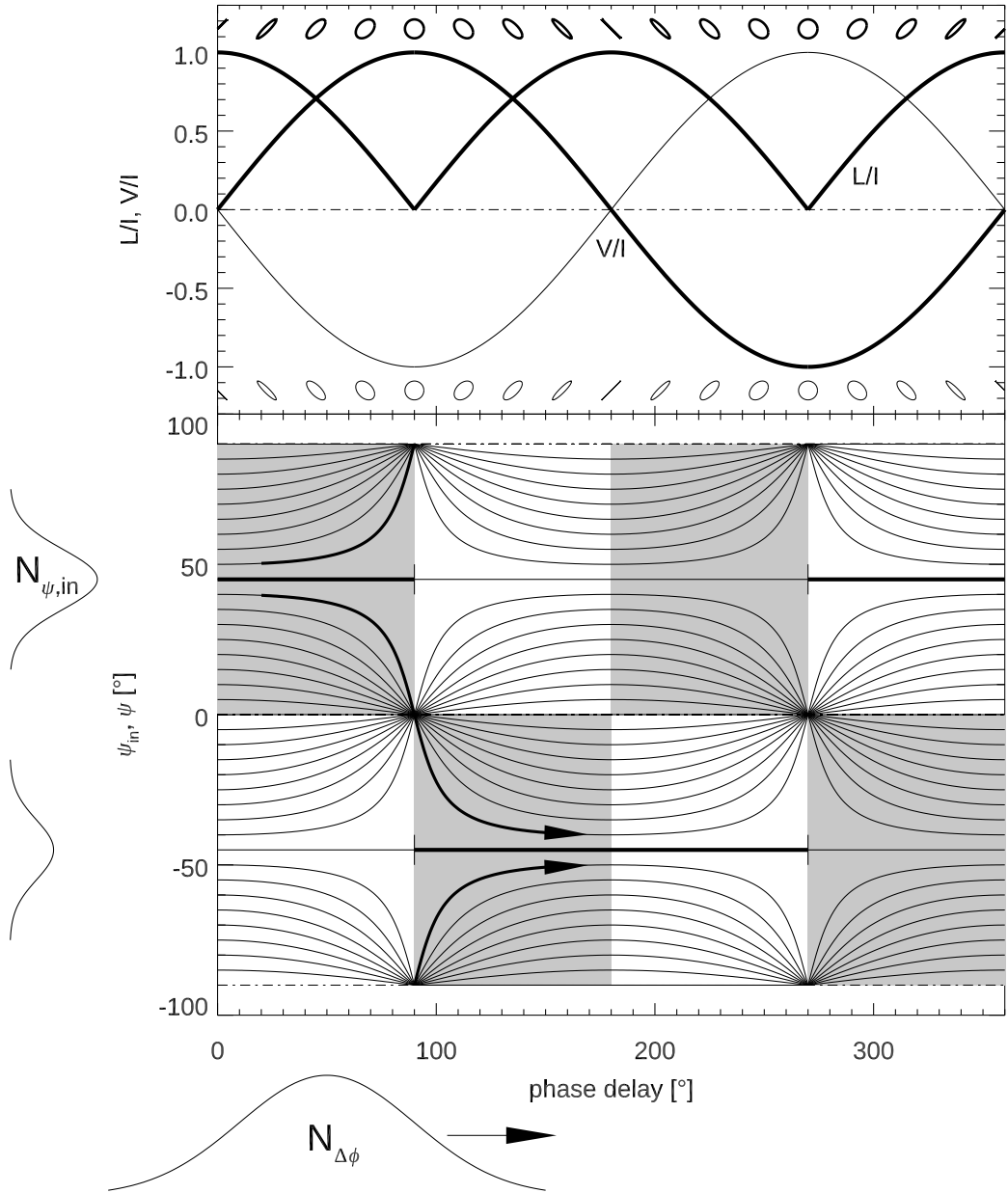}
\caption{Polarization characteristics of a signal that is a coherent sum of
two orthogonal linearly polarized waves. Bottom panel: PA as a function of 
phase lag. Different curves correspond to a fixed wave amplitude ratio
$E_2/E_1=\tan\psin$, with the mixing angle $\psin$ separated by $5^\circ$, as
shown on the vertical axis ($\psin=\psi(\lag=0)$).
The horizontal pieces of straight lines at $\pm45^\circ$ correspond to polarization ellipses
shown in the top panel. Grey rectangles present regions with positive $V$. 
The thick arrows diverging from
$\psin\approx45^\circ$, and converging at $\psin\approx-45^\circ$ present
the phenomenon of the lag-driven PA-track bifurcation. Statistical distributions of lag and mixing angle
are shown on the margins. The $\npsi$ distribution has two components (at
$+45^\circ$ and $-45^\circ$) which produce opposite
circular polarization. Top panel: polarization ellipses and polarized fractions $L/I$ and $V/I$ for
the equal amplitude case of $\psin=+45^\circ$ (thick solid) and $-45^\circ$ (thin). 
$L/I$ is identical 
in both cases.
}
\label{modes}
\end{figure*}

\subsection{Modelling hints} 

The PA loop of B1933$+$16 has been interpreted in D17 as a sudden rise (and
a following drop) of a phase lag between two linearly polarized orthogonal
waves, supposedly representing the natural propagation modes. The model is
quite successful because it can reproduce all relevant polarization characteristics,
such as the nearly bifurcated distortion of PA, the twin minima in $L/I$,
and the single-sign $V$ with a maximum at the modal
transition. Moreover, with a change of a single parameter (amplitude ratio of
modes) the model consistently reproduces the change of these features with frequency
$\nu$. All this occurs because within the loop, the underlying PA track (which
gets split into the loop as soon as the lag is increased) is assumed to be displaced by about 
$45^\circ$ from the linear natural modes.

However, the data (see Fig.~1 in MRA16) clearly show that the loop
opens on a PA track that can be considered as one of the normal OPMs (as
evidenced by a regular OPM jump observed just left to the loop). 
The model thus requires the modal power to be
$\sim\negthinspace\negthinspace45^\circ$ away from where the
power is actually observed to be, 
 if the identification of the observed OPMs as coincident with the normal modes is
correct. As described in D17, the observed OPMs are sharp spikes of
radiative power with the PA coincident with that of the natural propagation modes. 
As shown in Fig.~\ref{prinz} the observed orthogonal modes (M$_1$ and M$_2$) are produced when the phase lag 
distribution extends to $90^\circ$, since
at this value a linearly polarized input signal of any orientation is always
decomposed into 
polarization ellipses aligned with the linearly polarized natural (proper) modes
m$_1$ and m$_2$. As emphasized in D17, the observed OPMs (M$_1$ and M$_2$) are not the same as the proper modes 
m$_1$ and m$_2$, because M$_1$ and M$_2$ may have the same handedness despite being orthogonal
to each other 
(such case is shown in Fig.~\ref{prinz}). 
However, M$_1$ and M$_2$ have the same PA as the proper (normal) waves. Therefore, the
linearly-fed observed OPMs M$_1$ and M$_2$ will be called below the `coproper' modes.

Because the above-described $45^\circ$ difference was hard to justify, the analysis that followed
in D17 attempted to interpret the core polarization through changes of mode
amplitude ratio with pulse longitude (instead of the phase lag). The
amplitude ratio was parametrized by the mixing angle $\psin$, i.e. the angle at
which the emitted signal was separated into two linearly polarized natural
mode waves (Fig.~\ref{prinz}a).
Because of the partial geometrical symmetry of the problem, slow changes of $\psi_{\rm in}$ with
pulse longitude were essentially able to justify the core polarization behaviour of
B1237$+$25, at least for the upper branch of its bifurcated PA track.

However, PSR B1237$+$25 exhibits two different states of subpulse
modulation: the normal state (N) and the core-bright abnormal state (Ab). 
In the N state the core PA mostly follows the upper branch of the split PA
track, 
whereas the lower branch is brightest in the Ab state (cf. Figs.~1 and 6 of
SRM13). Despite the change of the branch, however, the
sign of $V$ remains the same in both cases. 
In the $\psin$-based model of D17 (section 4.7 therein) this was impossible to achieve, because the 
diverging branches of the bifurcated track were interpreted purely through 
departure of $\psi_{\rm in}$ from a natural mode in two opposite directions,
and the predicted sign of $V$ is different on both sides of the proper mode. 
This can be seen in the lag-PA diagram of Fig.~\ref{modes} 
which presents 
selected polarization properties of a wave that is a coherent combination of two
orthogonal and linearly polarized waves oscillating at a phase lag $\Delta\phi$. 
The sign of $V$, as represented by the grey and bright rectangles, is
opposite on each side of $\psi_{in} = 0^\circ$ (which corresponds to one
natural mode).
Moreover, both the loop of B1933$+$16 and
the PA bifurcation of B1237$+$25 look as phenomena of the same nature, so
it is not Ockham-economic to interpret them in different ways (change of
phase lag versus change of mixing angle). 
In the case of the PA loop of B1933$+$16, the model based on the $\psin$ only, 
could reproduce the twin $L/I$ minima, the single-sign $V$, and the PA
distortion, but was incapable to produce the loop-shaped bifurcation itself
(see Fig.~11 in D17). The bifurcation, instead, required the change of the
lag (Figs.~12 and 13 in D17). 

Below I further elaborate the models of
D17 in order to explain the mysterious $45^\circ$ misalignment which allows
us to
interpret both phenomena within a unified scheme.

\section{Introductory model}
\label{model}

\subsection{Coherent addition of linearly polarized waves}


Let us start with the model described in D17: before
reaching the observer, a radio signal of amplitude $E$ is decomposed into two linearly polarized waves with orthogonal
polarization:
\begin{equation}
E_x = E_1\cos{(\omega t)}, \ \ \ E_y = E_2\cos{(\omega t -
\Delta \phi)}.
\label{waves}
\end{equation}
The waves may be thought to represent the natural propagation modes of a
linearly polarizing, birefringent intervening medium. The main (proper)
polarization directions of the medium are $\vec x$ and $\vec y$.
After a phase delay $\Delta\phi$ is built up between the waves, they combine
(are added) coherently, which produces the detectable radio signal.  
The amplitudes of the combining waves are equal to
\begin{equation}
E_1=E\cos\psin{\rm,\ \ \ \ \ \ }E_2=E\sin\psin,
\label{amplits}
\end{equation}
where $\psin$ is the mixing angle that parametrizes the amplitudes' ratio:
\begin{equation}
\tan{\psin}=E_2/E_1=R.
\label{ratio}
\end{equation}
The Stokes parameters for the resulting wave (i.e.~calculated after the phase-lagged
components have been added coherently in the vector way) are given by:
\begin{eqnarray}
I  & = & E_1^2 + E_2^2 \label{sti}\\
Q & = & E_1^2 - E_2^2\label{qst}\\
U & = & 2 E_1 E_2 \cos(\Delta\phi)\label{ust}\\
V & = & -2 E_1 E_2 \sin(\Delta\phi)
\label{stokes}
\end{eqnarray}
whereas the linear polarization fraction and the resulting PA are:
\begin{eqnarray}
L/I &  =  & (Q^2 + U^2)^{1/2}/I\\
 \psi& = & 0.5\arctan{(U/Q)}.\label{psiang}
\label{psaj}
\end{eqnarray}
To calculate the observed PA, the coherent-origin angle of eq.~(\ref{psiang}) needs to be added
to the external reference value determined by the rotating vector model:
\begin{equation}
\psi_{\rm obs} = \psi + \psi_{\sss RVM}.
\label{totpsaj}
\end{equation}
Since we focus on coherent effects, only the value of $\psi$ will be
discussed below, but it must be remembered that $\psi=0$ corresponds to $\psi_{\rm obs} = \psi_{\sss
RVM}$, ie.~the RVM PA corresponds to the orientation of the intervening
basis vectors ($\vec x_1$ or $\vec x_2$) on the sky.

Diverse pairs of $(\lag,\psin)$ in such model give the polarization
characteristics presented in
Fig.~\ref{modes}. Different curves
in the lag-PA diagram (bottom panel of Fig.~\ref{modes}) present $\psi$ calculated for different values of
$\psin$. The value of $\psin$ is fixed along each line, except from the
horizontal lines at $\psin=\pm45^\circ$. In this equal-amplitude case the PA jumps
discontinuously by $90^\circ$, which corresponds to the transition of the
polarization ellipse through the circular stage (see the rows of ellipses in the top
panel). The grey rectangles (actually squares) represent the regions with
positive $V$. In spite of the impression made by the checkerboard pattern, 
the sign of $V$ can change only at $\Delta\phi=n 180^\circ$, where $n$ is an
integer. A change at $\Delta\phi=90^\circ+n 180^\circ$ is impossible, because
no lines cross these values of $\lag$, except at the dark nodes at 
corners of the grey regions. The nodes appear because for $\lag=90^\circ$ any orientation of the
incident wave polarization (hence any amplitude ratio $R$) produces a polarization ellipse
aligned with either $\vec x$ or $\vec y$ direction of the intervening
polarization basis (see D17 for more details). 

Because of the noisy nature of pulsar radio emission, in the following numerical
calculations the values of $\psin$ and $\lag$ are drawn from statistical
distributions $\npsi$ and $\nlag$ with peak positions $\ppk$ and $\lpk$ and widths $\sipsi$,
$\silag$. The intensity is taken as $I=\npsi(\psin)\nlag(\lag)$.
The results presented in sections (\ref{model})-(\ref{twofold}) are produced with the same numerical code which is
described in detail in sections $3.2.1$ and $3.2.2$ of D17. 

\begin{figure*}
\includegraphics[width=0.98\textwidth]{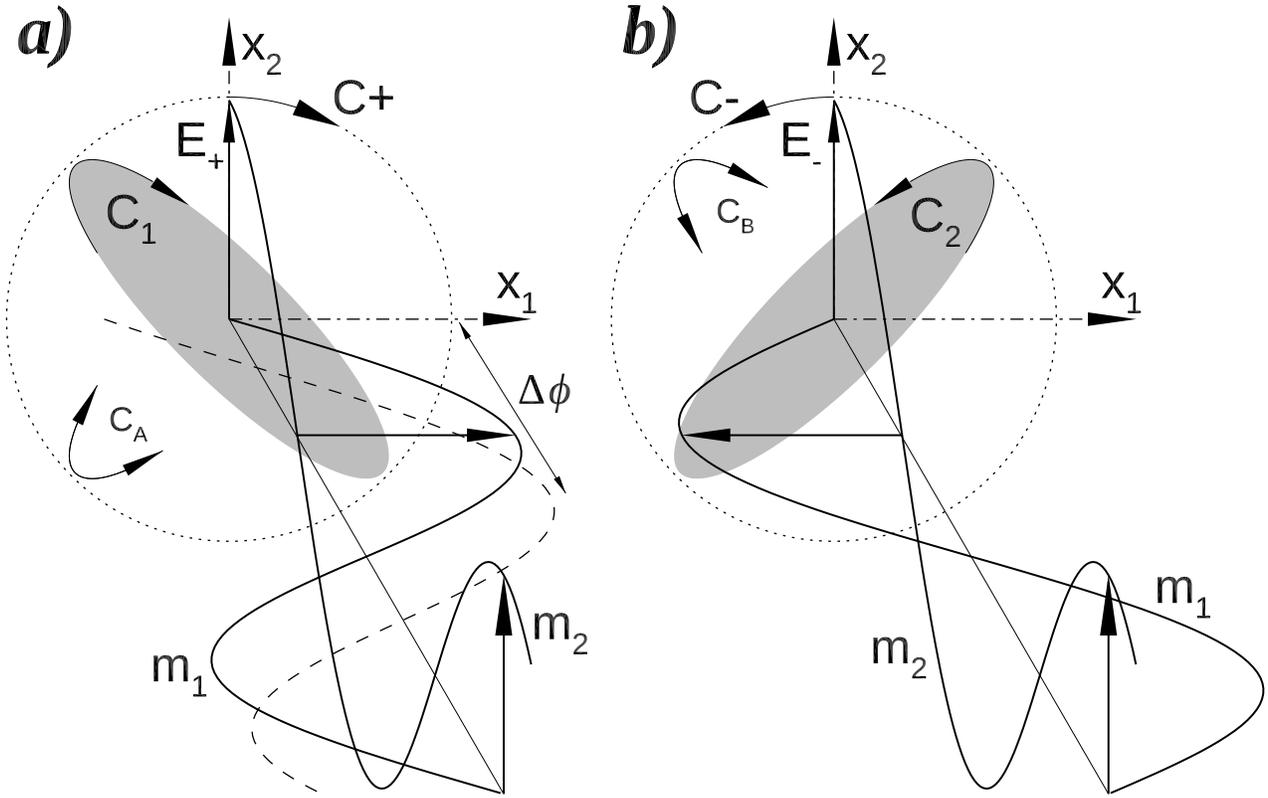}
\caption{The mechanism of generation of the observed orthogonal polarization modes
(grey ellipses C$_1$ and C$_2$) in the
case of equal amplitudes of the natural mode waves m$_1$ and m$_2$. 
While the proper mode waves m$_1$ and m$_2$ propagate through linearly
birefringent medium, they acquire a phase lag $\lag$ shown on the left. Then
they combine coherently into the grey ellipses of the observed OPMs, which are 
always tilted at $\pm45^\circ$ with respect to $\vec x_1$. Larger lags can
also produce the pseudomodal ellipses C$_A$ and C$_B$ of any handedness.
 The proper modal waves are fed by the circularly polarized signals C+ and
C-, which ensures equal amplitudes of m$_1$ and m$_2$. The circular feeding is not
essential for the model if the equal amplitudes are assumed ad hoc, but the
origin of the feeding circular waves may be elaborated to justify similar amounts of the observed OPMs 
C$_1$ and C$_2$ (Sect.~\ref{digression}).
}
\label{cirprinc}
\end{figure*}

\subsection{Another pair of orthogonal polarization modes -- equal wave
amplitudes}
\label{momo}

Unlike in D17, however, it is assumed in this section that the incident signal 
can be represented by two circularly polarized
waves of opposite handedness.\footnote{As discussed below, such circular
waves can be produced by a
decomposition of an elliptically polarized wave in medium with circularly
polarized natural propagation modes.} For simplicity of interpretation, 
in this introductory model the
detection of these  
circular waves (C+ and C- in Fig.~\ref{cirprinc}a and b) is assumed to be
non-simultaneous (ie.~the signal produced by C+ is not added coherently to
the signal produced by C-).   
Consider the wave electric vector $\vec E_+$ which traces
a spiral that projects on the dotted circle C+ in Fig.~\ref{cirprinc}a. 
In the aforementioned linearly-polarizing birefringent medium, the
wave induces the two linearly polarized waves, marked m$_1$ and m$_2$, and
described by eq.~\ref{waves}. The original phase delay $\lag$ between the waves is
equal to $90^\circ$ which results directly from their circular feeding.  
This phase lag $\lag$ is assumed to be increased (or decreased) by different refraction
indices of the natural propagation modes (therefore, the wave m$_1$ is shifted to
the dashed sinusoid position). Then the modal waves m$_1$ and m$_2$
are coherently added, which produces the 
elliptically polarized observed signal which is presented by the grey ellipse marked C$_1$. 
This is one of the observed OPMs (or one observed PA track, if the name OPMs
is to be reserved for the linearly fed coproper OPMs of D17). 
The eccentricity and handedness of the C$_1$ ellipse depends on the value of the
lag, however, as long as $\lag$ is between $90$ and $270^\circ$ all the resulting
ellipses will have the same PA, precisely at the angle of $-45^\circ$ with
respect to the PA of the natural propagation modes.\footnote{This makes such
circular-fed $45^\circ$-off modes statistically frequent, 
which is a feature analogical to the
linear-fed modes of D17.}
 Larger lags produce another 
orthogonal ellipse, which is marked C$_A$ in the figure. This second ellipse
is $90^\circ$ away from C$_1$ and, therefore, may possibly be called the other OPM. However, 
the C$_A$ mode may have the same handendess as C$_1$, so perhaps it
should be called a pseudomode.\footnote{Though C$_A$ may have the opposite handedness
too.} As explained below, to
account for the observed phenomenology, it is
necessary to introduce a separate circularly polarized signal of opposite
handedness, denoted with C- and $\vec E_{-}$ in Fig.~\ref{cirprinc}b. This
additional wave, in the same way as just described,  produces the second observed OPM, marked with
the grey ellipse C$_2$. Again, as explained below, the mode may be accompanied by
a pseudomode C$_B$, which may have the same or opposite handedness as C$_2$.

\begin{figure}
\includegraphics[width=0.49\textwidth]{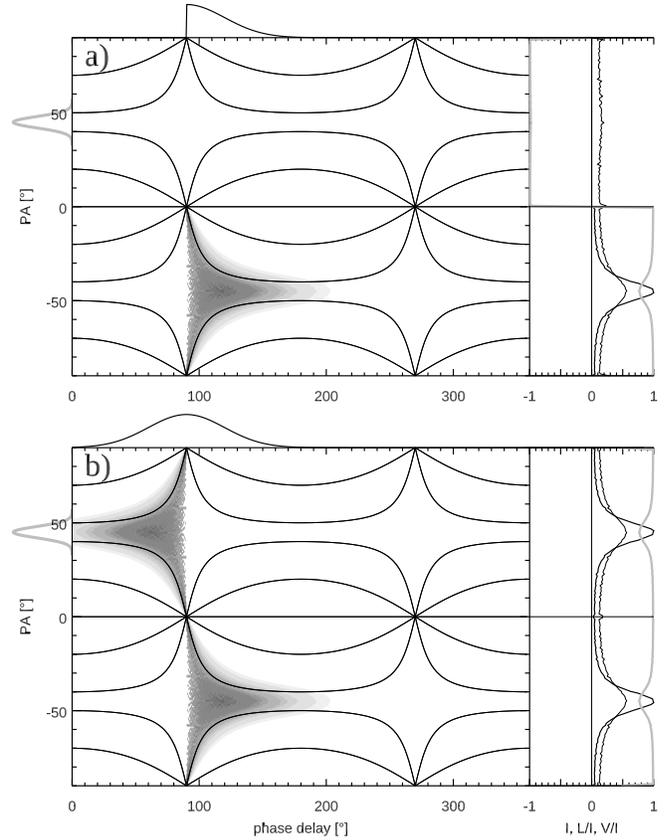}
\caption{Polarization characteristics for a mixture of non-simultaneous 
signals, each of which is composed of the coherently added natural modal waves 
m$_1$ and m$_2$. The phase lag $\lag$ and mixing angle $\psin$ for the
coherent addition were sampled from the statistical distributions shown near the top and left
axes of the plot (in {\bf a} $\nlag$ is asymmetric).  
The result can be considered to present distribution of many radio signal samples
observed at the same pulse longitude in different pulsar rotations. The main
panels present the pattern of observed radiative power on the lag-PA
diagram. The right panels present the observed intensity (thick
solid), $L/I$ (thin solid), and $V/I$ (light grey) as a function of
PA. Thus, the right panels show a vertical cut through the customary plots
of PA distributions that are
often presented for single pulse data. The result is for $\psin=45^\circ$,
$\sipsi=3^\circ$, $\lpk=90^\circ$ and $\silag=30^\circ$.
}
\label{twoplo}
\end{figure}

Remarkably, the new observed modes  C$_1$ and C$_2$ form a pair which is $45^\circ$ away from the natural
modes m$_1$ and m$_2$. They are also mid way between the linear-fed modes
described in D17, which have the same PA as m$_1$ and m$_2$.
The new circular-fed OPMs\footnote{The OPMs may also be called
same-amplitude OPMs, especially that the circular feeding may be considered
irrelevant to the problem as soon as the equal amplitudes are considered as an ad hoc assumption.
See, however, Sect.~\ref{digression}.} 
can be readily handled with the mathematical model described above,
because each observed mode results from coherent addition of phase lagged, linearly
polarized orthogonal waves (m$_1$ and m$_2$). 
Specifically, equal amplitudes of the waves imply the mixing angle of
$\pm45^\circ$ (eq.~\ref{ratio}) and the circulating feeding of the waves implies 
the initial phase lag of $\lag=90^\circ$. These are the positions at which the
waves $E_+$ and $E_-$ have to be injected into the lag-PA diagram of Fig.~\ref{modes}. 
Accordingly, Fig.~\ref{twoplo} presents the lag-PA pattern that appears for a single
feeding 
wave (C+) injected at $(\lag,\psin)=(90^\circ, 45^\circ)$. Each set of
panels in the figure may 
be considered as presentation of signals detected at a fixed pulse
longitude in many different pulse periods.  
The value of $\psin$ was sampled from a narrow Gaussian $\npsi$ distribution of
width $\sipsi=3^\circ$ 
whereas $\nlag$ had the width $\silag=30^\circ$ (both distributions are
shown near the plot axes).  
The right panels present the distribution of PA angles at a fixed pulse
longitude, i.e.~they present a vertical cut through those grey-scale PA
histograms that are usually shown for single-pulse data  
(the black thick solid line is the intensity cumulated at a given PA). 
The distribution of $V/I$ is shown with thick grey line and $L/I$ is thin
solid. 
Fig.~\ref{twoplo}a shows the case of a one-sided lag distribution, whereas the bottom
panels show the symmetric $\nlag$. Comparison of panels a and b implies 
that a single circular feed (eg.~C+ at $\psin=45^\circ$) can
produce two orthogonal PA tracks depending on the shape and position of
$\nlag$. The difference of refraction indices favours the one-sided $\nlag$, 
and it is also the case which avoids some depolarization
typical of the two-sided $\npsi$. 
For the moderately wide lag distribution 
used in Fig.~\ref{twoplo}, the power stays close to $\lag=90^\circ$ and therefore $V/I$ is high
($\sim\negthinspace\negthinspace0.7$). $L/I$ is about $0.5$ in the top case, and the same in both PA
tracks of the bottom-right panel. However, after Stokes-averaging over the PA distribution,
the average $L/I$ (at some longitude $\Phi$) would be very low, unlike in the top case.
The symmetry of $\nlag$ distribution is thus important for
some conclusions of this paper. 

As can be seen in Fig.~\ref{twoplo}, the lag distribution is extending the grey PA pattern
horizontally at $\psin=\pm45^\circ$ and it is these horizontal extensions
(which can look as dark horizontal bars -- see the next figure) 
that correspond to the observed OPM ellipses C$_1$ and C$_2$ in Fig.~\ref{cirprinc}. 
The more these `dark modal bars' are centered at $\lag=n180^\circ$, 
the higher
is the local $L/I$ (in a single PA track) and the smaller is $|V|/I$ 
(this can be deduced from top panel of Fig.~\ref{modes}).  

The important general implication of this section is that after statistical
averaging over $\nlag$, the observed OPMs (or the observed PA tracks) have the PA that is \emph{different} from the
PA of the natural mode waves m$_1$ and m$_2$ (this PA is equal to 0 or
$90^\circ$, as measured from the $\vec x_1$ axis of Fig.~\ref{cirprinc}). In the specific case
considered (equal amplitudes of the natural modes, $\psin=45^\circ$), the
observed OPMs are located mid way between the natural modes. Thus, \emph{the
observed PA tracks are not equivalent to the natural mode waves.} As shown further below
(Sect.~\ref{noneq}), 
the PA tracks may in general be displaced by an arbitrary,
mode-amplitude-ratio-dependent angle (and a $\nu$-dependent angle) from the natural modes. 
In the special case of equal amplitudes (of the natural mode waves m$_1$ and
m$_2$) the two observed PA tracks (C$_1$ and C$_2$, or C$_1$ and C$_A$) 
are separated by $90^\circ$ from each other, and can easily be misidentified as the natural
orthogonal modes, although they are misaligned by $45^\circ$ from the
natural modes m$_1$ and m$_2$.

\begin{figure}
\includegraphics[width=0.49\textwidth]{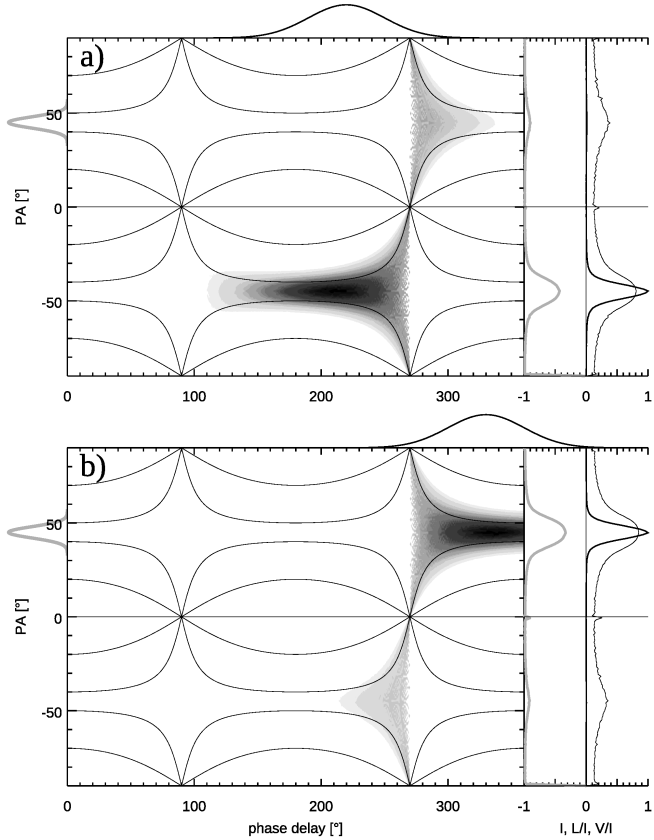}

\caption{The effect of increasing phase lag $\lag$. The $\nlag$ distribution is moving
rightward. In {\bf a} $\lag=220^\circ$ which causes the sign of $V$ in the
bottom PA track to change (compare $V$ in the bottom PA maximum of
Fig.~\ref{twoplo}b), but the PA maximum stays at $\psin=-45^\circ$. 
In {\bf b} $\lag$ increases to $330^\circ$ which moves the observed PA to $+45^\circ$, but
the sign of $V$ does not change. This is called a pseudomodal behaviour.
To obtain the normal OPM behavior, another $\npsi$ distribution would have to be added at
$-45^\circ$, and the relative power in both $\npsi$ would have to change
with pulse longitude. The observed modes have the form
of dark horizontal bars clearly visible in both lag-PA diagrams.
}
\label{lagmot}
\end{figure}

\section{Basic properties of the equal-amplitude OPMs and their application to
pulsar problems} 
\label{interp}

For the circular origin of the coherently combined waves m$_1$ and m$_2$, the value of $\ppk$ is
fixed and $\npsi$ must be narrow. Therefore, we are left with only three different processes 
that can happen to the radiative power on the lag-PA
diagram: 1) the lag distribution may move to larger (or smaller) values; 2)
the lag distribution may become wider, and 3) the other orthogonal circular-fed OPM
can be added as an additional $\psin$ distribution at $-45^\circ$, i.e.~the ratio of
amplitudes of $E_+$ and $E_-$ may change. 

\subsection{Movement of the phase lag distribution}

Fig.~\ref{lagmot} shows what happens when the phase lag distribution of
Fig.~\ref{twoplo}b moves towards larger values of $\lag$.
In Fig.~\ref{lagmot}a $\nlag$ is centered at $220^\circ$ which is larger
than $\pi$, hence $V$ of the bottom PA track (at $-45^\circ$) becomes negative 
(compare the grey curves in \ref{twoplo}b and \ref{lagmot}a). 
The sign of $V$ can thus change within the same PA track. 
In Fig.~\ref{lagmot}b the PA track makes an OPM transition to the upper
value of $+45^\circ$, however, the circular polarisation stays negative, as
in Fig.~\ref{lagmot}a.
Thus, the increase of the lag can cause some OPM transitions, but they
do not coincide with the sign change of $V$. They are actually a quarter of
lag-change cycle away, so that the OPM jump  occurs at a maximum $V$, while the sign
change of $V$ occurs well within a stable modal PA track, i.e.~within
flatter parts of a `non-transiting' observed PA track. This is similar to the
lag-driven effects in the coproper modes described in D17. 
The orthogonal modal tracks created 
by the change of
lag (or by widening of $\nlag$) can thus be called pseudomodes -- 
they do not obey the normal rule of zero $V$
at the minimum $L/I$. 
A phenomenon of this type (ie.~lag-change-based)  is observed in several pulsars, eg.~in the core PA
bifurcations. The lag-driven transfer of power between different OPM
tracks also explains the same sign of $V$ in different OPM tracks, as observed in 
single pulses (MAR15). 
Superficially similar pseudomodal behaviour is also observed in the form of slow OPM
transitions at high $|V|$ that occur within the whole pulse window 
(eg.~in PSR B1913$+$16, see Fig.~1 in D17, after Everett \& Weisberg 2001, 
also PSR J1900$-$2600, Johnston \& Kerr 2018). 
\nct{ew2001, jk2018} However, these are probably caused by the PA wandering which is
discussed further below. 

\begin{figure}
\includegraphics[width=0.49\textwidth]{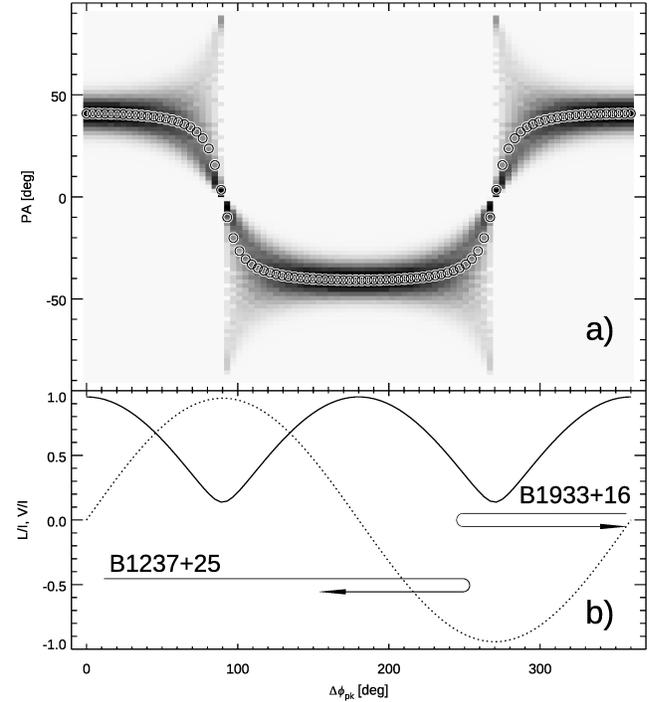}

\caption{Top: Lag-PA pattern calculated for
$\ppk=41^\circ$, $\sipsi=5^\circ$, $\silag=15^\circ$. Horizontal axis presents $\lpk$.  
Circles show the PA averaged over $\npsi$
 and $\nlag$ distributions.
The corresponding $L/I$ and $V/I$ is
shown in {\bf b}. A change of lag corresponds to the horizontal
motion in this plot, as shown with the backward bent arrows. 
Two interpretations of the core polarization, shown for B1237$+$25 and B1933$+$16,
work at a single frequency, but fail to explain the frequency behaviour of
the phenomenon.
}
\label{pecomp}
\end{figure}

The appearance of the equal-amplitude observed modes at the $45^\circ$ distance from the natural modes 
is interesting: it \emph{seems} to 
automatically solve the problem of what the primary observed OPM is doing half way between the
natural modes at
the entry to the PA loop of B1933$+$16. It is sufficient to claim 
that the observed OPMs are $45^\circ$ away from the natural modes, because their amplitude ratio is
close to $1$ at this particular frequency. 
In such case, the PA loop can be explained by a rise and drop of $|\lag|$, such as marked with the
backward-bent arrow in Fig.~\ref{pecomp}b (right). The resulting loop is
shown in Fig.~13 of D17. Such model reproduces several observed properties,
such as the bifurcation of the PA track, twin minima in $L/I$, and the
single-sign (negative) $V$. Moreover, a change of lag within a larger
interval, such as shown in Fig.~\ref{pecomp}b (left) would explain the PA
track bifurcation of B1237$+$25, along with the sign-changing $V$ at the
core component. 

The lag-induced PA bifurcation is also illustrated with
the curved arrows in Fig.~\ref{modes}. It can be seen that for the PA track
to split, the radiative power must be close to $\psin=45^\circ$ (in such
case the lines of fixed $\psin$ diverge up and down from $45^\circ$). 
Both the upward- and downward-heading arrows remain all the time within the
grey rectangles of positive $V$. Thus for the lag-induced
PA bifurcation the sign of $V$ stays the same whether the upper or bottom branch
of the bifurcation is followed. This would explain why the sign of $V$ is the
same in both modulation states in B1237$+$25: in the bright-core Ab
modulation state the lower branch of the bifurcation is followed, but the sign of $V$
does not change (in comparison to the N state). 

It is thus found that the lag change is the key factor that affects the
PA bifurcations observed both in B1933$+$16 and B1237$+$25. Both these
phenomena have the same nature, and can be explained by the same model with
slightly different parameters.  
However, the PA loop of B1933$+$16
may also be interpreted in a different (and better) way, which retains the usual
coproper OPMs
(with the same PA as the natural waves m$_1$ and m$_2$ at 0 and $90^\circ$), 
but assumes a quick change of $\psin$ towards
$\sim\negthinspace\negthinspace45^\circ$ within the loop. This 
new interpretation is favoured as discussed later, but such new model also requires the rise and
drop of the phase lag within the loop. 

The PA bifurcation model that is based purely on the
lag-change faces serious problems when
the loop of B1933$+$16 is interpreted at two frequencies. In Sect.~4.4.1 of
D17 (cf.~Figs.~13 and 14 therein) I have shown that a change of a single parameter -- $\psin$ -- from
$41^\circ$ (at $1.5$ GHz) to $31^\circ$ at $4.5$ GHz well reproduces the new
look of the loop at the higher frequency. This can be inferred from
Fig.~\ref{modes}: the curved arrow that follows $\psin=40^\circ$ produces
 the PA amplitude of almost $90^\circ$. A similar arrow (not shown) that
would follow $\psin=30^\circ$ for the same range of lag, would produce a smaller amplitude of
PA, consistent with the data at $4.5$ GHz (see Fig.~1 in MRA16). 
The problem is that the change of $\nu$ is most naturally associated with
the change of phase lag. Even if the mode amplitude ratio (hence $\psin$)
changes with $\nu$, it is hard to argue that the lag $\lag$ does not change. 
For a smaller $\lag$, the  
horizontal backward-bending arrow in Fig.~\ref{pecomp}b (right) would turn back earlier,
which would have made the PA amplitude smaller (as observed). However, such
earlier 
backward turn would also cause the twin minima in
$L/I$  to approach each other, or even merge into a single minimum at
the middle of the loop. This is not observed at $4.5$ GHz: the minima in
$L/I$ become very
shallow but stay at the same $\Phi$ (Fig.~1 in MRA16).

Apparently the lag-change alone cannot explain the loop at both frequencies. 
It will be shown below that simultaneous change of $\psin$ and $\lag$ with
pulse longitude is
needed to understand the phenomenon at both frequencies.

\subsection{Changes of width of the lag distribution} 
\label{randomization}

Considerable widening of the lag distribution wipes out the circular polarization and
tends to produce two highly linearly polarized PA tracks of similar or
equal amplitude (which gives zero net $L/I$ at a given $\Phi$). This is
because the radiative power is filling in several `dark horizontal bars' at
both $45^\circ$ and $-45^\circ$ in the lag-PA
diagram. On the other hand, for moderately strong widening of $\nlag$ the results may be similar to those
of the $\nlag$ shift, because the `center of weight' of the widening
$\nlag$ moves rightward.

\begin{figure}
\includegraphics[width=0.49\textwidth]{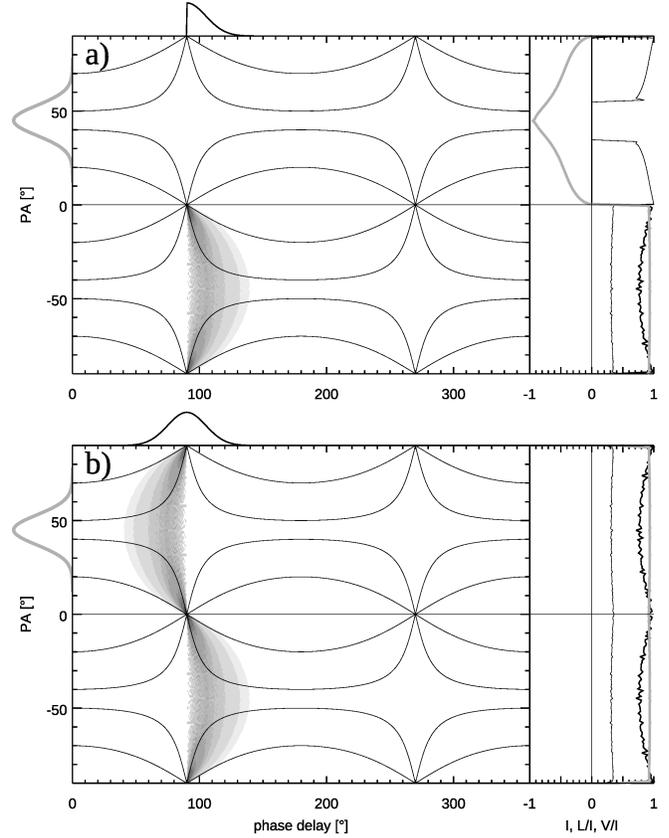}

\caption{The effect of $45^\circ$ PA jump, with the associated randomization
of PAs visible in the right panels. When $\nlag$ becomes narrow in
comparison to $\npsi$ the maxima of PA distribution move to $0$ and
$90^\circ$ (compare the PA distributions in Fig.~\ref{twoplo}). The small
differences of lag around $\lpk=90^\circ$ result in the randomisation of the
observed PA, which is similar to the behaviour of B1919$+$21 at 352 MHz
(MRA15).  
}
\label{forfi}
\end{figure}

An interesting effect appears when the lag distribution becomes narrow and
has a comparable width to the $\psin$ distribution ($\npsi$). Fig.~\ref{forfi} has been
calculated for $\sipsi=8^\circ$ and $\silag=15^\circ$. As can be seen in the
right-hand panels, this makes the PA distribution quasi uniform, and the peaks relocate to coincide with the
natural propagation modes m$_1$ and m$_2$ (located at PA of $0$ and $\pm90^\circ$). The degree to which the peaks
stand out depends on the ratio of $\sipsi$ and $\silag$,
and increases for narrower $\nlag$.
This phenomenon has therefore the key characteristics of the $45^\circ$ PA
jump, namely, the randomization of PA and the appearance of new pair of preferred
 PA values which are displaced by $45^\circ$ from the equal-amplitude OPM tracks (observed
in the wide-$\nlag$ case).

The modelled quasi uniform distribution of PA corresponds to the erratic PA
spread observed in the central profile region of PSR B1919$+$21, 
where the average PA curve is displaced by $45^\circ$  
 (see Fig.~18 in MAR15). The chaotic (quasi-uniform in the model) distribution
of PA becomes visible because the narrow $\nlag$ is negligible, so the
observed signal directly presents the state with no additional phase lag between the linear 
components m$_1$ and m$_2$. In this way the circulating motion of the electric
field $E_+$, as presented by the dotted circle C+ in
Fig.~\ref{cirprinc}a becomes directly visible (the circulation is recovered
as the sum of the m$_1$ and m$_2$ waves with the little-changed original phase delay
of $90^\circ$).
Surprisingly, then, according to the circular-fed equal-amplitude model, 
the observed erratic PA spread also has
geometric origin: it results from the circulating motion of the incident
circularly polarized signal. 
The observed $45^\circ$ PA jump thus represents 
the transition from the lag-spread-stabilized PA (which represents the state of
quasi-noncoherent average) to the lag-sensitive chaos of coherent states. In
such model, the narrow well defined PA tracks present the observed OPMs (ie.~the grey
ellipses C$_1$ and C$_2$ that are misaligned by $45^\circ$ from the natural
modes) which are associated with an average of wide
$\nlag$-distribution.  The longitudes with the erratic PA, on the other
hand, present the non-averaged emission in which case the natural
propagation modes m$_1$ and m$_2$ get through essentially undelayed.
This interpretation, therefore, also associates the observed OPM tracks
with the intermodes, just as the aforedescribed PA bifurcation model does. 

The PA randomization of Fig.~\ref{forfi} has been obtained for a single OPM
signal (say, C$_1$ fed by C+, contributing the $\npsi$ distribution at
$+45^\circ$ in Fig.~\ref{cirprinc}). 
In this case the circular polarization can stay larger than zero throughout the $45^\circ$
jump, as observed in B0823$+$26 (Everett \& Weisberg 2001) \nct{ew2001}
 The addition of the second orthogonal mode (C$_2$ or C- in
Fig.~\ref{cirprinc}) allows to suppress $V$ arbitrarily strongly.

The phenomenon of the $45^\circ$ jump was interpreted in D17 as the
narrowing of the lag distribution, which is maintained here. However, the
orthogonal modes that correspond to the wide $\nlag$, and are observed at
the profile outskirts in B1919$+$21,  were 
interpreted differently, and the peak of the $\npsi$ distribution 
in the narrow lag state was arbitrarily positioned near $\psin=45^\circ$.
In the model discussed in this section the nature of the observed OPMs is different (circular fed C$_1$ and
C$_2$) 
and they automatically tend to stay at the $45^\circ$ distance from the
orthogonal proper waves (m$_1$ and m$_2$).

The widening and displacements of $\nlag$ produce the psedomodal behaviour
-- 
they are incapable of reproducing the classical mode jumps with coincident
minima of $L/I$ and $|V|/I$. To obtain such regular behaviour it is necessary to
introduce the second circularly polarized component that produces the C$_2$
OPM. This rises the question of why the amplitudes of these circular waves tend to
be close to each other, and what causes the amplitude ratio to invert at the
regular OPM jumps.


\begin{figure}
\includegraphics[width=0.49\textwidth]{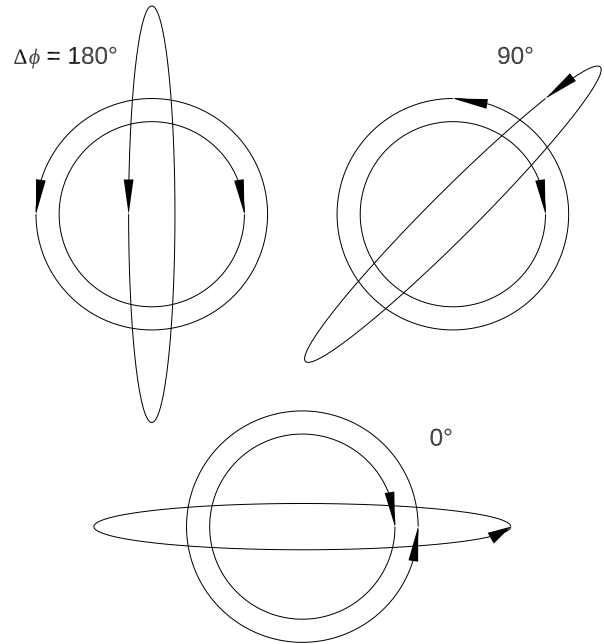}

\caption{The almost linearly polarized signal, represented by the narrow polarization
ellipse, can be mathematically decomposed into circularly polarized waves of similar
amplitude. The same decomposition can occur in circularly birefringent
medium. Rotation of the quasi-linear signal only changes the oscillation phases of the
circular waves (tips of arrows refer to the same moment of time, and numbers give the
phase diference). The waves' amplitudes are unaffected (cf.~the Faraday rotation
effect). 
}
\label{cireli3}
\end{figure}

\begin{figure}
\includegraphics[width=0.49\textwidth]{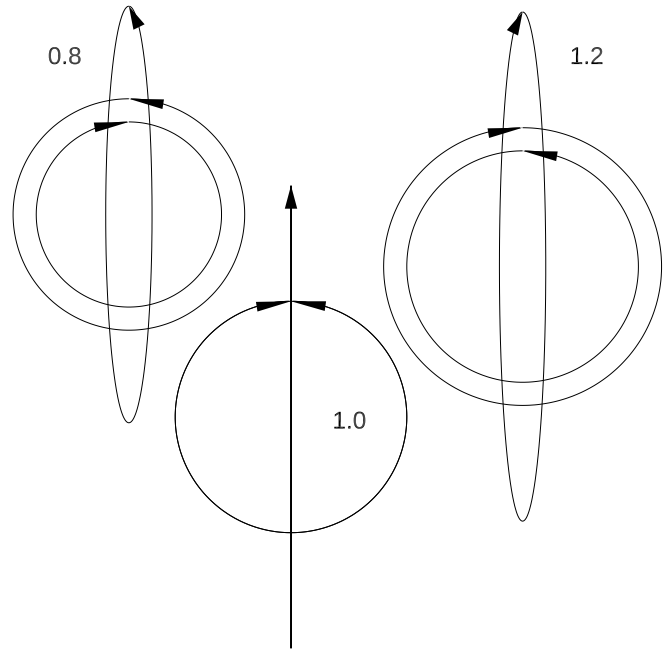}

\caption{Decomposition of the quasi-linearly polarized signal into circularly
polarized waves. The amplitude ratio of the waves (equal to $0.8$, $1$ and
$1.2$ in the cases shown) depends on the eccentricity of the signal's polarization
ellipse. With the change of handedness, the relative amplitude of the
circular waves is inverted.
}
\label{cireli4}
\end{figure}


%


\section{Pulsar as a pair of birefringent filters}
\label{digression}

\subsection{Similar amount of pulsar OPMs in the circular-fed model}

It has been shown above that some pulsar polarization effects can be
described as the linearly birefringent filtration of two circularly polarized waves of
similar amplitude but opposite handedness.  
If added coherently, such circular waves combine into a
linearly polarized wave, or an elliptically polarized wave with large 
eccentricity of its polarization ellipse (see Figs.~\ref{cireli3} and
\ref{cireli4}). This suggests that 
both these circularly polarized waves (C+ and C-) are generated by a single
emitted signal with a narrow polarization ellipse. Such original signal may be split into
two circularly-orthogonal waves in medium with circularly polarized natural
propagation modes. If the original (i.e.~emitted) signal is 
completely linearly polarized, as in the middle case in Fig.~\ref{cireli4}, 
then it produces identical amplitudes of both 
these circular waves (C+ and C-).

In D17  the circular wave stage of the model was absent. 
Along with the change of pulse longitude
$\Phi$,  the electric vector $\vec E$ of the emitted linearly polarized  signal was slowly rotating with respect to the
intervening polarization basis $(\vec x_1, \vec x_2$). The reason was the
change of angle between the low- and high-altitude direction of charge
trajectories (section 4.5 therein).
Whenever the vector $\vec E$ was passing through the $45^\circ$ angle, ie.~mid
way between $\vec x_1$ and $\vec x_2$, the mode amplitude ratio was
inverted. This was causing the $90^\circ$ OPM jumps, albeit of the pseudmodal nature 
(with $|V|$ peaking at the minimum $L/I$). 
 
In the present circular-fed model such effect is not possible, because the
rotation of the initial linearly polarized (or slightly elliptical) signal does not affect the
amplitudes of the circular waves C+ and C-. The rotation just changes the phases of
the waves, as shown in Fig.~\ref{cireli3} (it is the Faraday rotation
effect). Whatever the absolute oscillation phase 
of the circular waves,
they always feed the same, orthogonal and linearly polarized waves m$_1$ and m$_2$.
However, 
the ellipticity and handedness of the initial signal do affect the
amplitudes of C+ and C-.
As illustrated in Fig.~\ref{cireli4}, the relative amount of the circular waves
is inverted whenever the handedness is changed, and the amplitude ratio is
determined by the eccentricity of the initial signal.

Therefore, in the model with the filtration of the initial signal by the
circularly-birefringent medium 
the regular OPM jumps (with the coincident minima of $|V|/I$ and $L/I$) 
are caused by the change of handedness of the emitted signal. It is the handedness of the emitted radiation which determines which circle in
Fig.~\ref{cirprinc} is larger, and which $\npsi$ distribution -- whether the
one at $45^\circ$ or the one at $-45^\circ$ -- is stronger, i.e.~higher.

\subsection{The regular OPM jump in the center of radio pulsar profiles}

There is a way to test the hypothesis that the regular modal jumps are
caused by the handedness change. 
It is well known that the regular inversion of the mode amplitude ratio is often observed in
the central parts of pulsar profiles. The millisecond pulsar PSR
J0437$-$4715 provides an example of this effect, as evidenced by the sign change of
$V$ and OPM jump at the normally behaving $L/I$ minimum (see
Fig.~\ref{j0437}). 
Such sign-changing, sinusoid-like profile of $V$ has long been associated with a sightline traverse through 
a fan beam of curvature radiation, the latter being emitted by a bent stream of charges 
(e.g.~Michel 1991, pp.~355 - 359). 
\nct{m91}
The ensuing pulse of curvature radiation, at least
in vacuum theory, has precisely the sinusoid-like, handedness-changing
profile of $V$. As a consequence of the geometry shown in Figs.~\ref{cireli4}
and $\ref{cirprinc}$, there should be a regular $90^\circ$ OPM jump
produced by the change of handedness, and it is indeed often observed at zero $V$
in such core components of supposedly curvature-radiation-related origin.\footnote{The
orthogonal elliptically polarized modes have by definition
the opposite handedness, so it may seem to be a trivial vicious circle argument 
that a change of $V$ sign confirms a modal jump. However, it is not, 
because without the final linearly birefringent filtering, the circular waves of
Fig.~\ref{cireli4} would combine back to the original ellipse or would be
observed as separate circularly polarized signals. So it is the pair of
filters 
which produces the regular OPM jumps.
}

Thus, pulsar magnetosphere consisting of two filters that are made of circularly and
linearly birefringent materials provides a quite successful polarization model: it
is capable of explaining both the above-described non-RVM peculiarities and the standard
polarization properties such as comparable modal power and the regular OPM jumps. 
However, such model is complex and difficult to justify physically. 
A possible physical scenario would include a low altitude emission of the
nearly linear signal, followed by the circular decomposition in weak magnetic
field at large altitudes. The final stage of the linear filtering could
possibly be considered as equivalent to the effects that occur at the polarization limiting radius.   
Because of  this complexity, in what follows the relative
amplitude of the opposite-$V$ modes is considered as a free parameter.
 
\section{Twofold nature of pulsar polarization}
\label{twofold}

\begin{figure}
\includegraphics[width=0.49\textwidth]{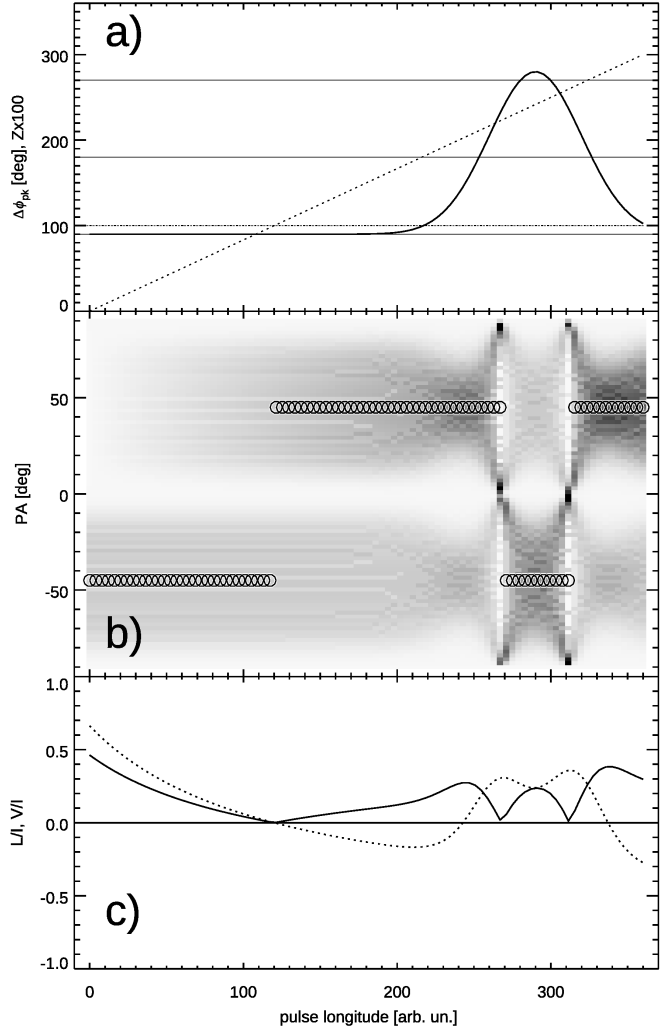}

\caption{Two generic polarization effects in the coherent wave addition
model: the regular OPM jump caused by inversion of mode amplitude ratio
(left, at $\Phi\approx120^\circ$) and the lag-driven PA bifurcation/loop effect (right,
at $\Phi\approx290^\circ$), both shown
as functions of pulse longitude. Top: the
relative power Z of both modes (dotted diagonal shows $Z\times 100$) 
and the function
$\lpk(\Phi)$ which follows the Gaussian centered at
$\Phi=290^\circ$. Beyond the Gaussian $\lpk$ is equal to $90^\circ$. The profiles of $L/I$
(solid) and $V/I$ (dotted) are shown at the bottom. The lag-profile Gaussian has
the $1\sigma$ width of $30^\circ$, whereas $\sipsi=13^\circ$, $\silag=45^\circ$.
}
\label{arti2}
\end{figure}
\begin{figure}
\includegraphics[width=0.49\textwidth]{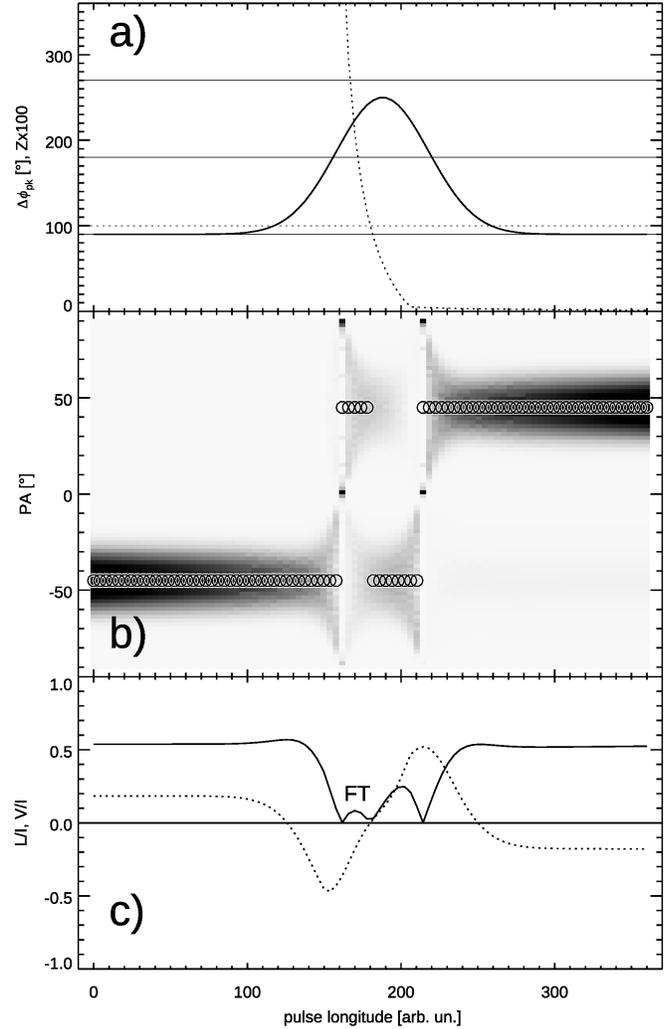}

\caption{The result of simultaneous operation of both effects -- the lag
change and amplitude change -- within the same longitude interval. Both the
$Z=1$ crossing point and the Gaussian lag-pulse are moved toward the center
of the $\Phi$ axis. Note the appearance of dissimilar twin minima denoted
`FT' that coincide with OPM jumps. The left minimum has high $|V|/I$, the
right one coincides with $V=0$. This is the phenomenon observed in the
profile center of J0437$-$4715. Since $Z$ and the amplitude of the
lag pulse depend on $\nu$, the look of this
effect strongly depends on frequency. The value of $Z$ is changing as shown
with the dotted line in {\bf a}, $\sipsi=5^\circ$, $\silag=105^\circ$.
}
\label{frat}
\end{figure}

It was shown above that aside from the RVM effect, the polarization of pulsar radio signal can change because of
two independent reasons: 1) as a result of change of the lag distribution
$\nlag$ and 2) as a result of change of the modal amplitude ratio (expressed 
by the ratio $|\vec E_+|/|\vec E_-|$ in the circular-fed model). The first factor likely depends on
the local properties of the intervening matter: a temporary increase of
refraction index may appear when the line of sight is traversing through some extra
amount of matter, e.g.~a plasma stream. The second factor is likely governed by the
radio emission process (and is determined by the ellipticity and handedness of the
emitted radiation in the specific case of the filter pair model). 
The two mechanisms -- the lag-driven and amplitude-driven
changes of polarization -- have
markedly different properties. The lag-driven effect produces the
anticorellated variations of $|V|/I$ and $L/I$ with pulse longitude (and OPM
jumps at maximum $|V|/I$). The amplitude-driven effects generate the regular
modal behaviour with the usual OPM jumps.

These generic properties are illustrated in Fig.~\ref{arti2} which presents
a regular OPM transition on the left ($\Phi\approx120^\circ$) and the
lag-driven 
bifurcation of the PA track on the right ($\Phi\approx290^\circ$) as a function of
pulse longitude. The regular OPM
coincides with the mode amplitude ratio of $1$. The relative power of both
modes, hereafter denoted $Z$, can be expressed as the integrated
power (or just height, in case of identical width) of the $\npsi$ distribution
at $+45^\circ$ and $-45^\circ$. The increasing value of $Z\times 100$ is shown in the
top panel (dotted), along with a temporary increase of $\lpk$
(solid Gaussian).\footnote{Here `temporary' means `constrained to a narrow interval
of pulse longitude'.} Several polarization effects observed in
radio pulsars result from either process, or from a mixture of both.
As described in section \ref{obs}, both these non-RVM effects appear to shape the
observed polarization especially in the central parts of pulsar profile.

Both these effects may depend on frequency.
The influence of the lag may depend on $\nu$ because the lag depends on the refraction index, which is likely
$\nu$-dependent.
As for the amplitude-related effects, they need to be $\nu$-dependent to
explain the modal power exchange observed in the D-type pulsars by Young
and Rankin (2012).
\nct{yr12}
This exchange of power seems to coincide with the $V$-sign
change, although the radio spectral coverage is far from
continuous.\footnote{In the case of B0301$+$19, for example, between the 327 MHz and
$1.4$ GHz Arecibo profiles of Young \& Rankin I have only found the $1.22$ GHz profile in
MRA15. The modal exchange takes place near $1.22$ GHz since $V$ has both
signs in different parts of the average profile at this $\nu$.}
 The $\nu$-dependent amplitude ratio is also responsible for
 another type of PA distortions (slow PA wandering) that is discussed further below.


\subsection{The origin of dissimilar $L/I$ minima in PSR J0437$-$4715}

While interpreting polarization in the central (or other) 
parts of any profile it is important  to allow for the possibility that both the effects of
lag and amplitude ratio may be overlapping there to produce a net profile of
$L/I$, $V/I$ and a net PA. An obvious example of such overlap is
the center of the profile of J0437$-$4715.

Fig.~\ref{frat} presents model result for the case when the temporary rise
of lag (solid line in top panel) roughly coincides in $\Phi$ with the amplitude ratio reversal
(dotted curve in top panel). The parameters have been changed a bit
in comparison to Fig.~\ref{arti2}, eg.~the rate of $Z$ change was increased, 
however, the main difference is that the
longitude of equal mode power ($Z=1$) and the peak of $\lpk(\Phi)$ profile were displaced to
roughly the same $\Phi$. 
This combination of lag and amplitude effects reproduces the major features of the
central profile portion in J0437$-$4715 at 660 MHz (Fig.~\ref{j0437}, after
NMSKB97).
A double minimum of $L/I$ appears at $\Phi=170^\circ$ (denoted FT in
Fig.~\ref{frat}). The
right minimum in this pair coincides with the change of $V$ sign, whereas the left one 
coincides with high $|V|/I$. The value of $L/I$ in the regular right minimum does not quite reach
zero, as in the observation. Within the longitude interval flanked by
the minima, the PA is visiting the orthogonal PA track, but quickly returns
back to the $-45^\circ$ value.\footnote{Since the model $\npsi$ is perfectly
aligned with $45^\circ$, one is free to choose whether the PA jump direction is up or
down.} 
The deep minimum at $\Phi\approx215^\circ$
does not follow the observations, but this is only because no efforts have
been made to adjust parameters in this longitude interval. 
Another difference is that the modelled OPM follows the full $90^\circ$
traverse. This is caused by the perfect alignment of the $\npsi$ with
$\pm45^\circ$ (this constraint will be relieved below). 

The complex polarization of core emission can thus be understood as a
combination of the lag-driven and amplitude-ratio-driven polarization effects. 
The core emission of normal pulsars (eg.~B1237$+$25, B1933$+$16) also
exhibits polarization profiles that are neither symmetric nor antisymmetric. 
Apparently, the overlap of lag and amplitude effects also occurs in these
objects and is partially destroying the anti/symmetry of $L/I$ and $V/I$
which appears when the lag and amplitude phenomena are viewed separately.

\subsection{Towards a general model}
\label{towards}

Let us summarize the results obtained so far.
A model based on coherent and quasi-coherent addition of linearly polarized
waves of roughly equal amplitude is capable of qualitatively reproducing polarized
profiles (ie.~all three components: $L/I$, $V/I$ and PA) of the following phenomena:
1) the bifurcations of PA track in pulsars with complicated core emission 
(ie.~B1933$+$16 and B1237$+$25, including two modulation states of the
latter) and 2) the mixed core behaviour of J0437$-$4715. 
When extended to encompass the origin of the feeding circular
waves, the model can possibly justify the regular OPM jumps and
the similar amount of modes. 

On the other hand, the purely linear birefringent filtering may seem unphysical, and
the model faces two problems that contain indications about how 
to change it. First, the pseudomodal OPM transitions 
tend to traverse regions of very low $L/I$. 
As can be seen in Fig.~\ref{modes}, for $\lag$ increasing from zero 
at $\psin\sim45^\circ$ the radiative power approaches 
the fully circularly polarized point
at $(\lag,\psin)=(90^\circ,45^\circ)$ then jumps down to $\psi=-45^\circ$ 
while staying all the time fully circularly polarized. This is consistent
with the low $L/I$ observed at the core PA bifurcations in PSR B1933$+$16 and
B1237$+$25, however, a capability to flexibly adjust the modelled $L/I$ is
needed: in the D17's $\psin$-based model of
B1913$+$16 it was difficult to avoid the strong decrease of $L/I$ at the
OPM transitions (cf.~Figs.~1b and 7b in D17).
Second, with the circular feeding of the linear proper waves (m$_1$ and
m$_2$), the $\npsi$ distribution  is absolutely tied to
$\pm45^\circ$. Actually, even the spread of $\npsi$ around these values
(parametrized by $\sipsi$) is hard to explain.\footnote{The displacement 
from $\psin=45^\circ$ could be obtained for elliptically polarized feeding
waves.}  

The circularly polarized waves (C+ and C-) that feed m$_1$ and m$_2$ are
then too restrictive for the model and, at least when the `filter pair' concept is
dismissed, they indeed do little more than set the
equal amplitude ratio of m$_1$ and m$_2$. 
Therefore, in the following I will use the lone pair of standard, 
ellipitically polarized, orthogonal natural mode waves (EPONM waves). 
Obviously, the coherent addition of such waves \emph{must} produce all the
successful results of previous sections, because the linearly polarized
equal amplitude waves are just a special case of EPONM waves. 
However, the arbitrary amplitude ratio and the nonzero ellipticity provide 
important enlargement of the model capabilities. 

  
A general model of pulsar polarization thus includes the eccentricity of
the polarization ellipse for modal waves (m$_1$ and m$_2$). The
eccentricity parameter may need to be sampled from statistical distribution of some
width. Even with the same eccentricity for both modal
waves, this means two new parameters. Along with the other four (the mixing angle
for the amplitude ratio and the phase lag, plus the widths of their distributions), this makes up for six
parameters.  Such parameter space deserves a separate study, therefore,
in what follows I describe my calculation method and only present a glimpse of the
parameter space -- just to address the above-described problems. 

\section{General model}

\subsection{Coherent addition of elliptically-polarized orthogonal waves}
\label{general}

\begin{figure}
\includegraphics[width=0.49\textwidth]{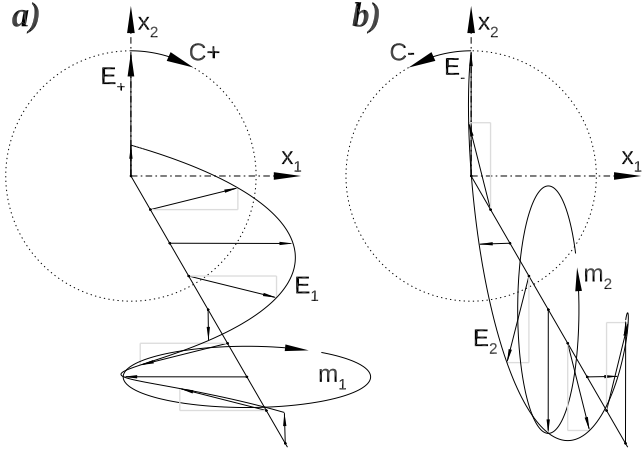}

\caption{General model of pulsar polarization is based on coherent addition
of elliptically polarized natural mode waves $E_1$ and $E_2$ which trace 
the solid line ellipses  m$_1$ and m$_2$  (shown in {\bf a} and {\bf b}). Here the
proper modal waves are fed by the circular waves $E_+$ and $E_-$ (see
Sect.~\ref{digression}), but this
is not necessary for the model to work.
}
\label{newp}
\end{figure}

\begin{figure*}
\includegraphics[width=0.89\textwidth]{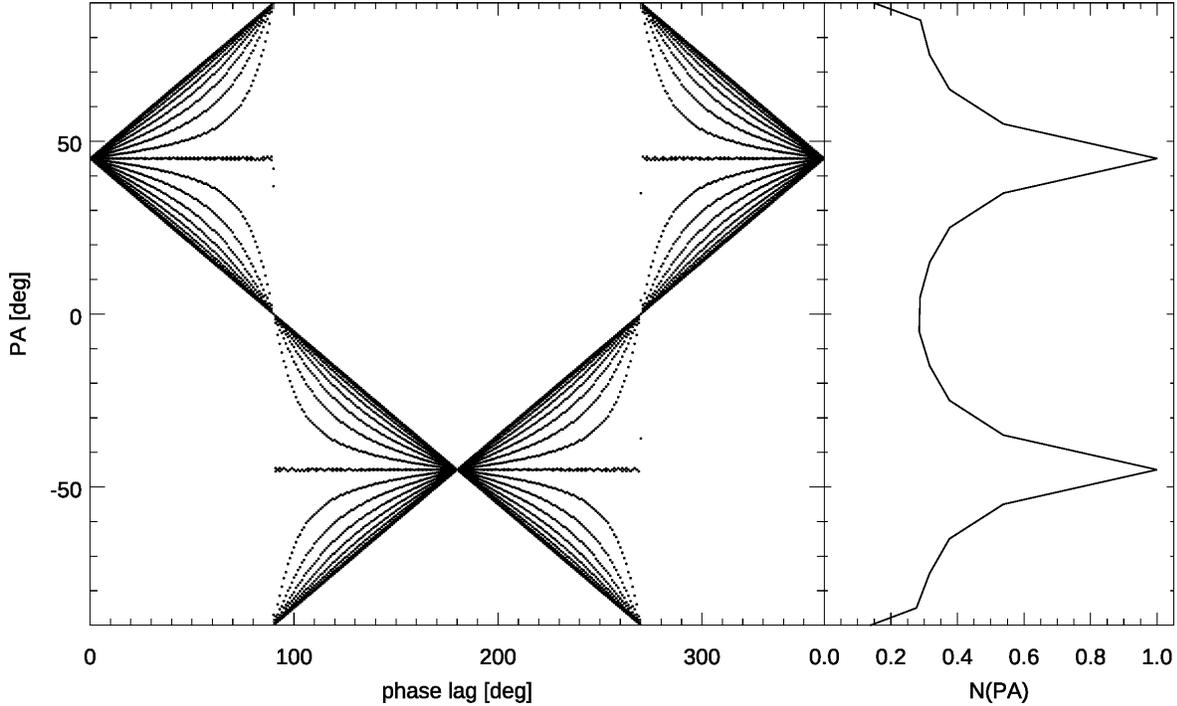}

\caption{Left: Lag-PA diagram for equal mode amplitudes ($\psin=45^\circ$) and for a set of eccentricity parameter
$\beta\in(0,180^\circ)$ sampled at interval of $5^\circ$. The horizontal
sections ($\beta=0$, and $90^\circ$) are produced by the linearly polarized proper modes
as described earlier
in this paper. The diagonals correspond to circularly polarized m$_1$ and
m$_2$ modes
($\beta=45^\circ$ and $135^\circ$) and are caused by the uniform PA change of
the resulting fully linearly polarized signal (close cases are shown in
Figs.~\ref{cireli4} and \ref{full2b}). There are new nodes at
$\lag=n180^\circ$ that coincide with the `dark modal bars' of the previous
analysis.  
Projection of all the PA angles on the vertical axis
gives the histogram shown on the right side.  The observed OPMs for
such equal-amplitude mixture of eccentricities are misaligned by $45^\circ$ from
the natural modes.
}
\label{full3}
\end{figure*}

\begin{figure*}
\includegraphics[width=0.89\textwidth]{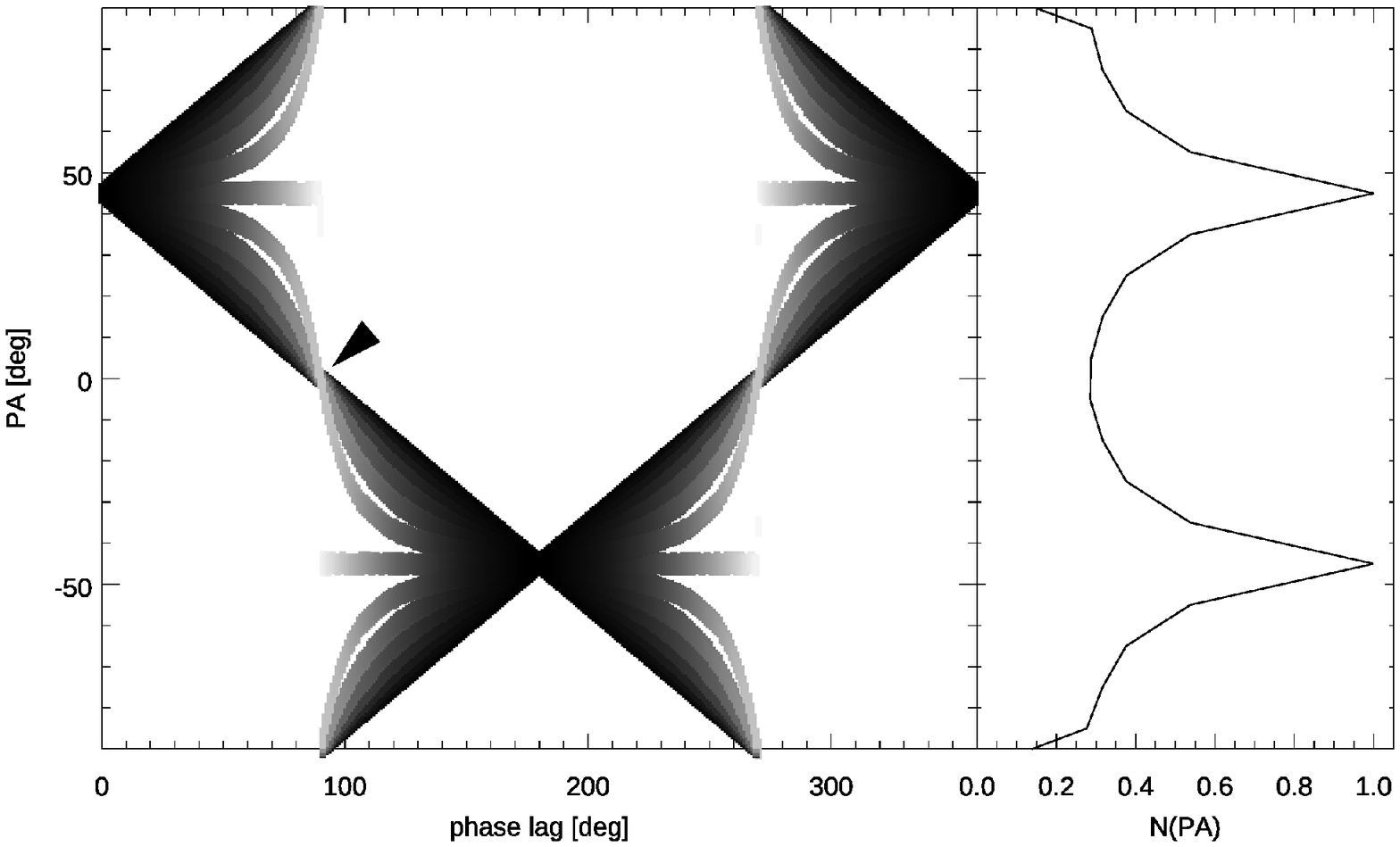}
\caption{Linear polarization fraction $L/I$ for the lag-PA diagram shown in the
previous figure. In the black regions $L/I=1$, whereas in the bright regions
$L/I=0$. At the point indicated by the arrow several cases of eccentricity
overlap, contributing different $L/I$. The linear (or quasi-linear) modes
($\beta=0$) contribute circular polarization ($L/I\sim0$) whereas the 
diagonals contribute linearly polarized signal ($L/I=1$). The pattern may
also be considered to present $|V|/I$, with the value of $1$ in the bright
regions, but $V/I$ changes sign several times in a given position of the
plot and would require a different type of visualization.
}
\label{full4}
\end{figure*}

The model is conceptually simple: observed pulsar polarization
results from coherent and quasi-coherent addition of phase-lagged waves in 
two elliptically polarized natural propagation modes. They are numbered $1$ and $2$
and are presented in Fig.~\ref{newp} by the ellipses m$_1$ and m$_2$. These
ellipses are traced by the corresponding electric field waves $\vec E_1$ and
$\vec E_2$. 
The ellipses m$_1$ and m$_2$ should not be mistaken
for the observed PA tracks, because the latter result from coherent addition of 
m$_1$ and m$_2$ and may be easily displaced from the natural modes by 
an arbitrary angle.
For example, if equal
amplitudes of m$_1$ and m$_2$ are preferred, then the observed polarization ellipses
(similar to the grey ellipses of Fig.~\ref{modes}) appear at the PA of
$45^\circ$ when m$_1$ and m$_2$ are coherently added.
The elliptical natural mode waves m$_1$ and m$_2$ can be written as:
\begin{eqnarray}
E_1^x&=&\cos\mix \cos\beta \cos(\omega t)\\
E_1^y&=&\cos\mix \sin\beta \sin(\omega t)
\end{eqnarray}
\begin{eqnarray}
E_2^x&=&\sin\mix \sin\beta \sin(\omega t + \lag)\\
E_2^y&=&\sin\mix \cos\beta \cos(\omega t+\lag)
\end{eqnarray}
where $\lag$ is the phase delay and  
$\tan\beta$ represents the ratio of the minor to major axis of the 
polarization ellipse. 
As usual $\mix$ represents the ratio of the modal waves' amplitudes, 
ie.~$\tan{\mix}=E_2/E_1$,
where $E_1^2=(E_1^x)^2+(E_1^y)^2$ and $E_2^2=(E_2^x)^2+(E_2^y)^2$.
These waves coherently combine into the observed signal that in general is elliptically
polarized:
\begin{eqnarray}
E^x&=&E_1^x+E_2^x\\
E^y&=&E_2^y+E_2^y
\end{eqnarray}
The polarization ellipse for the observed signal $\vec E=(E^x,E^y)$ is
calculated by numerically increasing $\omega t$ in the range between 0 and
$360^\circ$. The minor half axis $A_{\rm min}$ and the major half axis $A_{\rm max}$
of the observed ellipse are then identified numerically, along with the sense of the
electric vector circulation (handedness).  
The PA is determined by the normalized components of the major axis: 
\begin{equation}
\cos\psi=A_{\rm max}^y/A_{\rm max}, \ \ \sin\psi=A_{\rm max}^x/A_{\rm max}, 
\label{newpa}
\end{equation}
whereas the ellipse axes length ratio  gives the observed eccentricity angle:
\begin{equation}
\tan\beta_{\rm t}=A_{\rm min}/A_{\rm max}
\label{beta}
\end{equation}
which is different than the initial $\beta$ of the proper modal
waves. 
The normalized Stokes parameters are calculated from: 
\begin{eqnarray}
Q/I & = & \cos(2\beta_{\rm t})\cos(2\psi)\\
U/I & = & \cos(2\beta_{\rm t})\sin(2\psi)\\
V/I & = & \sin(2\beta_{\rm t}),
\end{eqnarray}
and the linear polarization fraction is calculated as
$L/I=[(Q/I)^2+(U/I)^2]^{0.5}$.

\subsection{Lag-PA diagrams for elliptical modes}
\label{lagpa}

The lag-PA diagram of Fig.~\ref{full3} (left panel) presents the pattern of PA calculated for
fixed values of eccentricity $\beta$ increasing uniformly from 0 to $180^\circ$
in step of $5^\circ$. The amplitudes of the combined modes are everywhere
the same ($\mix=45^\circ$).
The corresponding $L/I$ is shown in Fig.~\ref{full4}, with $L/I$ increasing
in darker regions. The sign of $V/I$ is changed several times in the same
points of this diagram and, therefore, $V/I$ is not shown. However, $|V|/I$ is as
before 
anticorellated with $L/I$, so dark regions in Fig.~\ref{full4} present low $|V|/I$.

The pattern presents new nodes, ie.~regions where there is high probablity to observe the
radiative flux. The nodes are at $(\lag,\psi)=(0,45^\circ)$ 
and $(180^\circ,-45^\circ)$.  
They appear for two reasons. The first is that for any eccentricity, at $\lag=n 180^\circ$ the
electric vectors of the equal-amplitude modal waves always combine at the PA that is $45^\circ$ 
away from the PA of the proper
modes (m$_1$ and m$_2$ have the PA of $0$ and $90^\circ$). The second reason
is that for the purely linear polarization (infinite eccentricity) of equal-amplitude
waves the resulting PA is equal to $\pm45^\circ$ regardless of $\lag$. This produces
the discontinuous PA jumps between the fixed PA values at $\lag=90^\circ$
and $270^\circ$. For high eccentricity (nearly linear modal waves m$_1$ and
m$_2$), and always for equal amplitude, 
the PA tends to linger close to $45^\circ$, which increases the probability of the
nearly intermodal PA. This is illustrated in Fig.~\ref{full2a} of
the appendix.

The model described earlier in this paper (with the equal-amplitude linearly polarized
orthogonal waves, LPOW)
was confined only to the horizontal PA segments centered at the new
nodes (and some nearby regions because $\npsi$ was allowed to have finite width). 
The lag-PA space of the LPOW model is just a subpart of the new lag-PA diagram and
this is because diverse ellipticities are added in Fig.~\ref{full3}. The
patterns of $L/I$ and $V/I$, within the overlapping part of the parameter
space,  are identical to the one of the LPOW model. 
For example, the 
value of $L/I$ in Fig.~\ref{full4} increases towards $\lag=n180^\circ$ and decreases at the
discontinous lag-change-driven OPM jumps. The new nodes coincide with the `dark
modal bars' of the LPOW model.    
This implies that all data interpretations provided before are also possible in
the new elliptical model. In other words, the added diversity of eccentricities does not 
corrupt the previous results.

When the PA values of the left panel are collected in bins on the vertical axis, the
histogram shown on the right is produced. The enhancements of the observed OPMs remain at the
intermodal positions (half way between the PA of the natural modes).
Naturally, for $\mix =45^\circ$ ($R=1$) the
intermodal nature of OPMs persists in the presence of diverse ellipticity.

A new feature of the lag-PA diagram are diagonal straight lines which
connect the nodes. These correspond to the sum of two circular waves
($\beta=45^\circ$) at
increasing lag. The result is a uniformly rotating linearly polarized
signal, hence the linear change of PA (the diagonals thus represent the
Faraday rotation effect).  
A similar case is shown in Fig.~\ref{cireli3} and Fig.~\ref{full2b} in the appendix.  
The linear polarization is full along the diagonals ($L/I=1$, $V/I=0$). 

If the radiation at a given pulse longitude contains a mixture of
eccentriticies, then the lag-driven OPM transition occurs both along the S-shaped (or discontinuous) 
paths in the lag-PA diagram and along the straight diagonals. As
shown in Fig.~\ref{full4}, in the middle of the OPM jump the combining
signals of high
eccentricity (ie.~the almost linearly polarized modal waves which follow the
S-shaped path) contribute circularly polarized power 
(note the bright stripe of the high $|V|/I$ at the position indicated by the
arrow) whereas the low-eccentricity signals (circularly polarized modal
waves that follow the diagonals) contribute linearly polarized power (the diagonals are black everywhere,
ie.~they have $L/I=1$, see Fig.~\ref{full2b}, compare Fig.~\ref{full2a}). 
The lag-driven OPM transition for a signal of mixed ellipticity, can thus be percieved
as the passage from, say, the top horizontal row in Figs.~\ref{full2b}
and \ref{full2a}, to the fourth row in these figures (along with all unshown
cases of intermediate ellipticity). As shown in Fig.~\ref{full4} with the
arrow, the inclusion of wider ellipses increases $L/I$ at the lag-driven OPM
jump. The inclusion of eccentricity can thus increase the very low $L/I$ at 
some lag-driven 
OPM transitions. 

\begin{figure}
\includegraphics[width=0.49\textwidth]{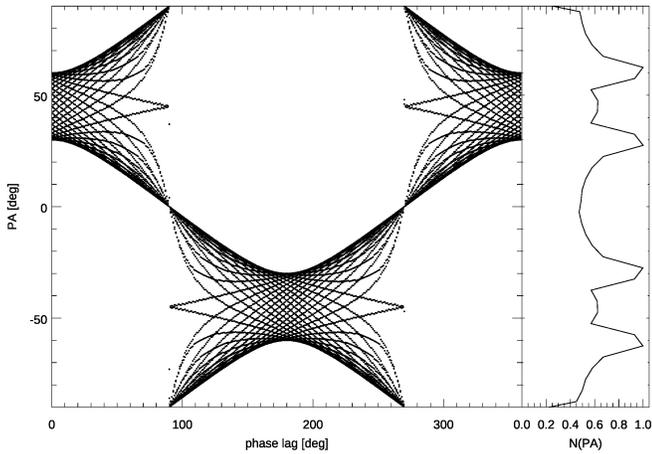}

\caption{The lag PA diagram for mode amplitude ratio of $0.58$
($\mix=30^\circ$). In comparison to the equal amplitude case, the curves of $\psi(\phi)$ have different shape 
and are displaced vertically. There is a spread of available PA values near the
nodes. Note the appearance of the non-orthogonal PA tracks in the histogram.
Not all peaks must be observed, depending on the actual spread of $\beta$
in the observed signal. 
}
\label{ampli2}
\end{figure}

\begin{figure}
\includegraphics[width=0.49\textwidth]{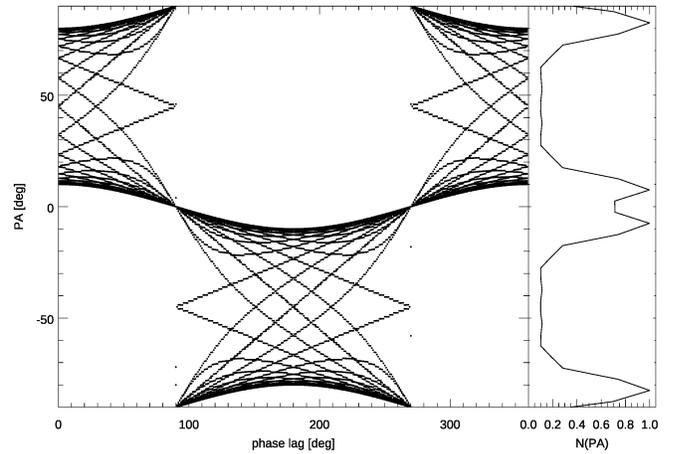}
\caption{
The lag-PA diagram for mode amplitude ratio of $0.18$ ($\mix=10^\circ$).
}
\label{ampli22}
\end{figure}

\begin{figure*}
\includegraphics[width=0.89\textwidth]{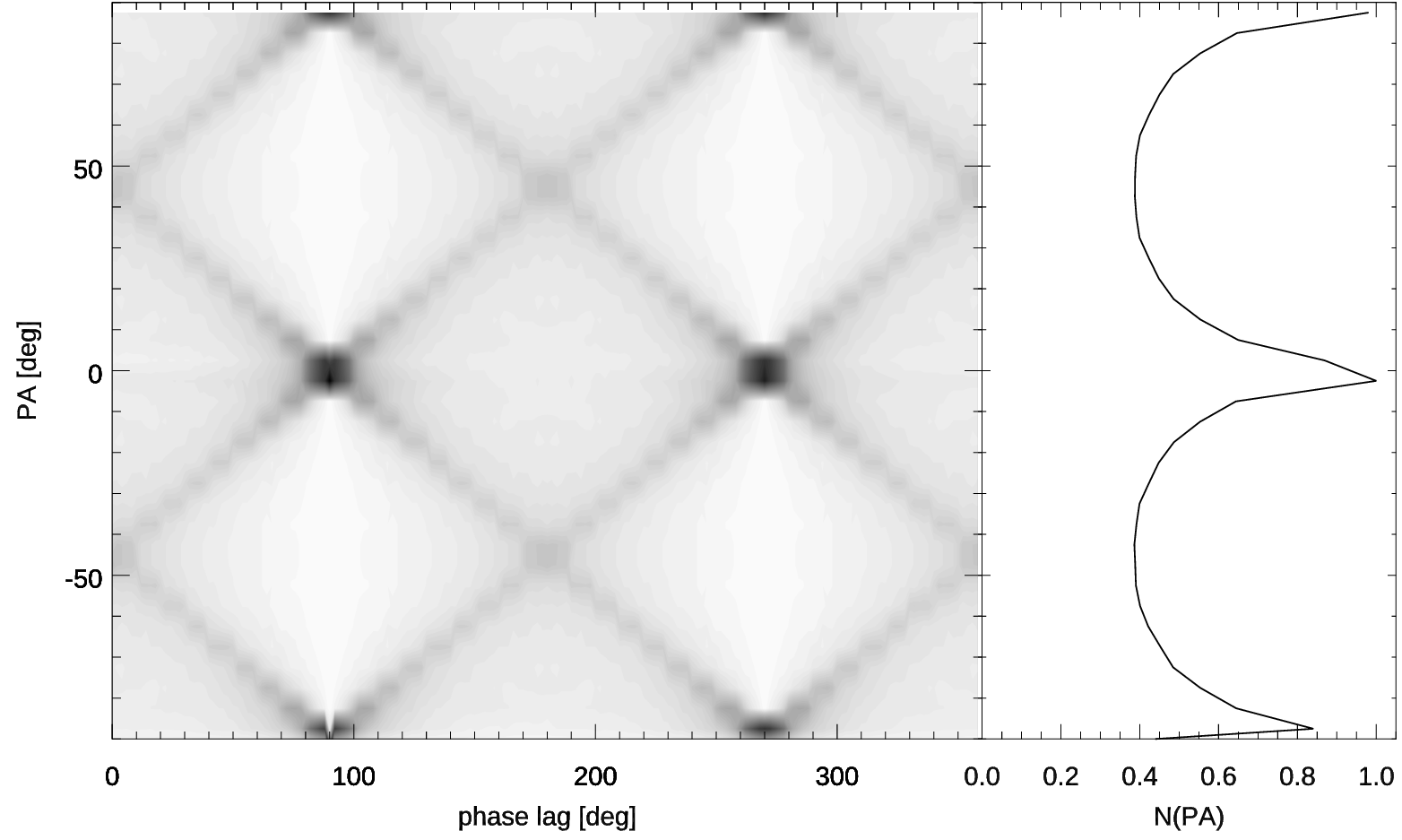}

\caption{The lag-PA pattern for densely sampled parameter space of $(\lag, \mix,
\beta)$. The sampling was uniform in all these three angles.
The observed PA tracks (peaks in the right histogram) coincide with the PA of the proper
mode waves m$_1$ and m$_2$. The black spots at the diagonals' crossings are at the same locations as the
linear-fed nodes described in D17 (also see Fig.~\ref{modes} in this paper).
}
\label{allp}
\end{figure*}

\subsection{Entanglement of the lag-driven and amplitude-ratio-driven
effects}

The amplitude ratio of observed OPMs seems to change with pulse longitude
$\Phi$ and with the frequency $\nu$. The lag change should be
considered as the primary effect which governs different look of PA tracks
at different $\nu$. The change of mode ratio (or $\psin$) is naturally
responsible for changes of polarization with $\Phi$ (as proved by the
regular OPM jumps). However, several observations at different $\nu$ suggest
that $\psin$ may also be $\nu$-dependent. Moreover, 
if some observed OPMs have the intermodal nature, as illustrated in
Fig.~\ref{cirprinc}, then it is the change of lag itself, which causes the
ratio of observed OPMs to change. This has been presented in 
Fig.~\ref{lagmot}, where the change of lag causes
the radiative power to leak from one orthogonal PA track to another. 
It should be possible to recognize if the observed change of OPM amount 
has the lag-driven origin, because the lag-driven effects exhibit the pseudomodal 
behaviour (anticorellation of $L/I$ and $|V|/I$). 
This complexity needs to be kept in mind when the non-equal mode amplitudes
are considered.

\subsection{Beyond the equal amplitudes}

To move $\npsi$ away from $45^\circ$, the amplitude ratio
of the natural mode waves (m$_1$ and m$_2$) must be changed to a less trivial
value than $1$.\footnote{To detach from $45^\circ$ in the
circular-fed equal-amplitude model, it is necessary to consider simultaneous detection
(and coherent combination) of
the modes C$_1$ and C$_2$.}
The change of $\mix$ causes the entire lag-PA pattern to evolve.
Fig.~\ref{ampli2} shows the case of $\mix=30^\circ$ (amplitude ratio of
$0.58$) whereas Fig.~\ref{ampli22} is for $\mix=10^\circ$ (ratio $0.17$).
As can be seen in Fig.~\ref{ampli2}, with the change of mode amount ratio
the PA paths move away from $45^\circ$. Moreover, with the increase
of lag many paths cover smaller range of PA than in the equal amplitude case. 


It should be noted that the addition of EPONM waves itself does
not imply any preference for the same or similar amount of modes. The intermodal 
observed OPMs (located at $\pm45^\circ$, see the histogram in
Fig.~\ref{full3}) are just the consequence of the 
equal amplitude assumption. 
When the entire parameter space is sampled uniformly in $\mix\in(0,180^\circ)$,
$\beta\in(0,180^\circ)$ and $\lag\in(0,360^\circ)$, the coproper modes of
D17 (with the same PA as the natural modes, see Fig.~\ref{prinz}) become statistically most probable and stand
out in the histogram (see Fig.~\ref{allp}). 

In the case of the linear-fed coproper modes (Fig.~\ref{prinz}), 
the equal amplitudes of observed OPMs are produced when the incident linear signal
is traversing through the intermodal separatrix IM (at a wide $\nlag$ that 
extends to
$\lag=90^\circ$). This seems to be a quite simple and natural way to change
the PA by $90^\circ$, but the sign of $V$ does not change along with the PA jump.
In the birefringent filter pair model, the equal/similar amplitudes of both
opposite-$V$ modes are produced by decomposition of the initial quasi-linearly
polarized signal in the circularly polarizing medium of the first filter. Apparently, some mechanism 
akin to such process is required to explain the nearly equal amount of the modes
m$_1$ and m$_2$ which are shown in Fig.~\ref{newp}.
In any case, if there is any reason for which the natural mode
amplitudes tend to be similar, then the histogram of Fig.~\ref{full3}
tells us that the observed OPMs are actually $45^\circ$ away from the
natural propagation modes (as illustrated in
Figs.~\ref{cirprinc} and \ref{full2a}). The dotted circles in Fig.~\ref{newp}
present how the similar amplitudes of modes can be produced if they are fed by the
circularly-polarized waves of common origin (ie.~produced by decomposition of an
almost linearly polarized signal into the C+ and C- waves of nearly equal
amplitude).

Finally, however, it must be admitted that amounts of the opposite-$V$ modes are equally often observed to
be strongly nonequal. This option opens the possibility to understand the
off-RVM wandering of PA tracks.

\subsection{Mode-intermode PA transitions and the origin of the 
off-RVM wandering of PA}
\label{noneq}

\begin{figure}
\includegraphics[width=0.49\textwidth]{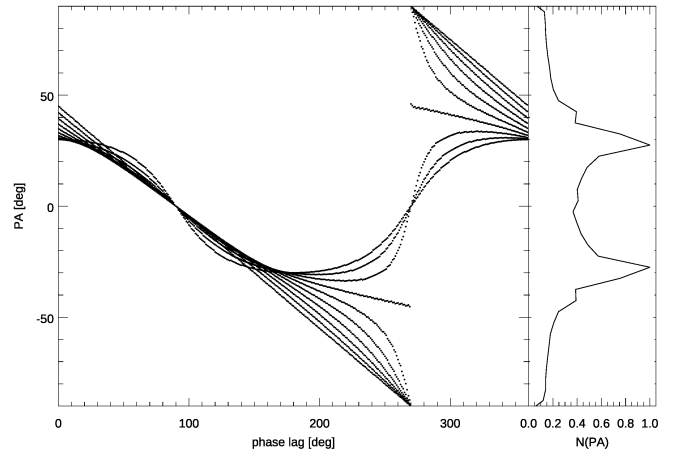}
\caption{
The lag-PA diagram for mode amplitude ratio of $0.58$ ($\mix=30^\circ$), but
with $\beta$ limited to the interval $(0,45^\circ)$. The resulting PAs are a subset
of 
those in Fig.~\ref{ampli2}. Two non-orthogonal PA tracks (visible as the
peaks in the right histogram) appear at
$\pm30^\circ$. Changes of $\mix$ would change the separation between the
tracks.
}
\label{amplidots}
\end{figure}

\subsubsection{Non-orthogonal PA tracks}

\begin{figure}
\includegraphics[width=0.49\textwidth]{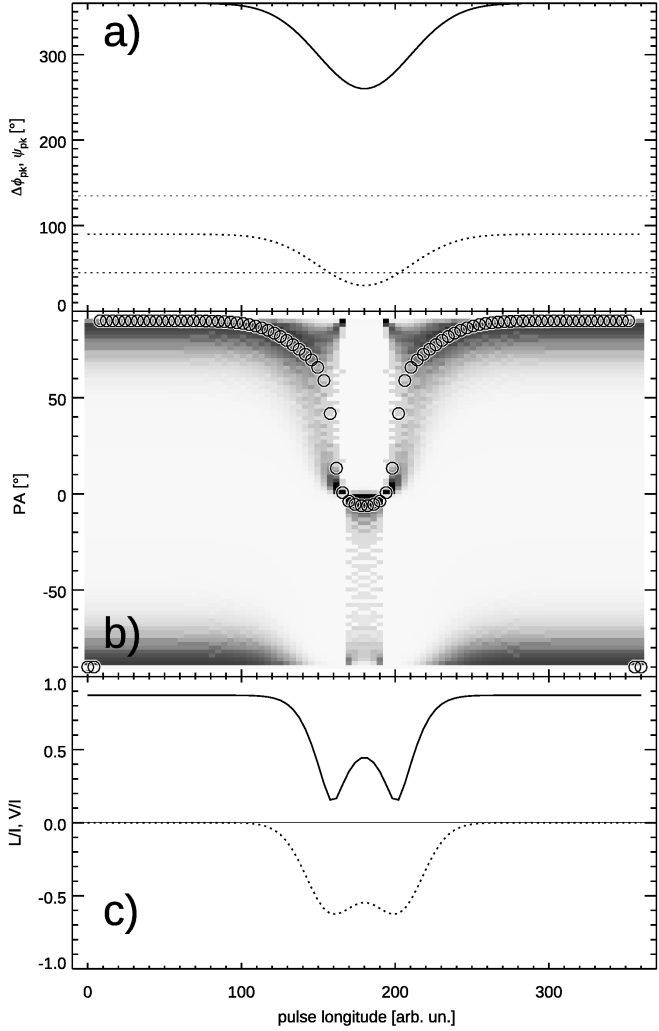}
\caption{
Modelled polarization characteristics for the PA loop of B1933$+$16 as
observed at $1.5$ GHz (cf.~Fig 1 in
MRA16). The loop shaped bifurcation of PA results from the Gaussian
variations of $\ppk$ and $\lpk$ (dotted and solid line in {\bf a}).
Distribution widths were kept constant across the feature: $\sipsi=15^\circ$,
$\silag=45^\circ$.  
}
\label{pal}
\end{figure}
\begin{figure}
\includegraphics[width=0.49\textwidth]{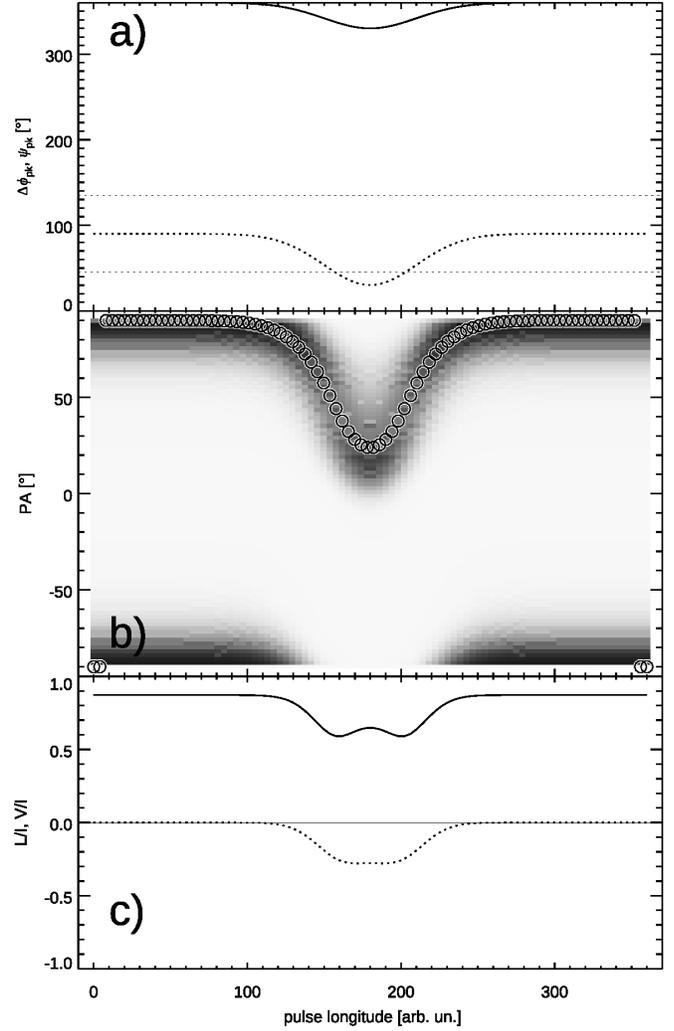}
\caption{Same as in the previous figure, but for a smaller amplitude of the
lag deviation (see the solid line in {\bf a}). The PA loop has been transformed  
into the U-shaped PA distortion similar to that observed in B1933$+$16 at
$4.5$ GHz. $L/I$ and $V/I$ also change in the way consistent with the data.
}
\label{palb}
\end{figure}

The primary mode exchange phenomenon, illustrated in figures of Young \&
Rankin (2012) and other observations clearly show that the mode amount ratio
changes with frequency. The ratio $Z$ (or $R$) also changes with pulse longitude
$\Phi$ as proved by the regular OPM jumps and ubiquitous changes of $L/I$ with $\Phi$. 

The coherent addition of proper modal waves (m$_1$ and m$_2$) implies that the
observed PA should be at $\psi=45^\circ$ (with respect to the proper wave
PA) whenever the lag distribution is wide and the modes' amplitudes are 
equal.\footnote{For a narrow $\nlag$ the observed PA becomes determined by
an accidental value of $\lag$ or $\lpk$. A wider $\nlag$ allows for the
statistically average value to stand out in a stable way.}  
Note that such statistical average of signals with different $\lag$
reproduces the PA value that would be expected for a coherent sum of linearly
polarized modal waves at zero lag,
ie.:
\begin{equation}
\psi=\psin=\arctan(E_2/E_1). 
\label{psinon}
\end{equation}
This equation is valid for other amplitude ratio values, however, for 
the full uniform distribution of $\beta$ the PA maxima appear as two pairs,
visible in the right panels of Figs.~\ref{ampli2} and \ref{ampli22},   
ie.~they appear at $\psi=\pm\psin$, and $\psi=90^\circ\pm\psin$. For a narrower
distribution of eccentricities only some of these four peaks remain in the
signal. Fig.~\ref{amplidots} is calculated for uniform $\beta$ in the range
$(0,45^\circ)$ and $\mix=30^\circ$ (hence this case is a subpart of 
Fig.~\ref{ampli2}). This gives two PA tracks at $\pm\psin$,
ie.~separated by $60^\circ$. The phenomenon of non-orthogonal PA tracks is
often observed in radio pulsar data. For example, in PSR B1944$+$17 and
PSR B2016$+$28 (Fig.~15 in MAR15) the PA tracks are neither orthogonal nor
parallel, which requires the nonequal modal amplitude to change with $\Phi$.

\subsubsection{Off-RVM PA wandering}

For increasingly non-equal amount of modes, $\psi$ will diverge from $45^\circ$ and, in
the limit of one mode absent, the observed PA must become equal to $0$ or
$90^\circ$, ie.~it must start to coincide with one of the natural modes. 
Therefore, the model predicts that whenever the mode ratio is changing
(along with
changing $\Phi$ or $\nu$) between $1$ and $0$, the observed PA should exhibit 
transitions between the observed intermodal PA track (at $\psi=45^\circ$) and the
proper mode at $\psi=0$. In a general case, the PA can wander between
some initial $\psin^{\rm init}$ and a final $\psin^{\rm fnl}$, with the values 
determined by their corresponding mode amplitude ratios, as given by
eq.~(\ref{psinon}).

This type of phenomenon likely occurs on
the right side of the 660 MHz profile in J0437$-$4715 (Fig.~\ref{j0437}).
The $\nu$-dependent modal ratio may also affect the strongly
$\nu$-dependent look of polarization within the core component of this
pulsar. However, figures \ref{full3}-\ref{amplidots} clearly show that changes of lag
with $\nu$ can affect the observed polarization at least equally strongly, and in
fact they do, as is shown in the following section.


\subsection{Frequency dependent lag and the loop of B1933$+$16}
\label{paloop}

The OPMs observed to the left of the PA loop in PSR B1933$+$16 are stable at
different frequencies: they look as the same pair of orthogonal patches at
both $1.5$ and $4.5$ GHz (see Fig.~1 in MRA16). This holds despite the
ratio of observed OPMs quickly changes with pulse longitude $\Phi$ at both
frequencies.
The steady orthogonal location of the modes can be ensured both in the
linear-fed model of Fig.~\ref{prinz} and in the circular-fed model of
Fig.~\ref{cirprinc}. In the case of the coproper modes of D17, the lag
distribution must be wide so that the nodes at PA of $0$ and $90^\circ$ 
are enhanced. For a narrow $\lag$ distribution the proper modal waves m$_1$
and m$_2$ would coherently combine to an arbitrary PA, as given by
eq.~(\ref{psinon}). Alternatively, the
proper modal waves would have to be detected non-simultaneously (to avoid
coherent addition) to hold the steady PA at $0$ and $90^\circ$. 

The observed minima in $L/I$ of B1933$+$16 (Fig.~1 in MRA16) reveal that the mode
amount ratio is being inverted  
every $4^\circ$ or so in the profile, so it is natural to assume that $R$
temporarily becomes close to unity ($\psin=45^\circ$ or $135^\circ$) within the loop
 (where `temporarily' again means `within the narrow interval of
$\Phi$'). 
Therefore, it is assumed in the following that $\psin$ slightly crosses the
value of $45^\circ$ within the PA loop. For simplicity, the eccentricity
is set to infinity (linear waves, $\beta=0$) so that the model considered is 
that of Fig.~\ref{prinz} (which is a special case of the general model shown
in Fig.~\ref{newp}). Moreover, since changes of $\lag$ appear indispensable
to obtain the bifurcation effect, it is assumed that both $\psin$ and $\lag$
change within the loop. 

Fig.~\ref{pal} presents polarization characteristics calculated for the
Gaussian change of $\ppk$ and $\lpk$, as shown in panel {\bf a} with the 
dotted and dashed line, respectively. Both Gaussians have the pulse
longitude width $\sigma=30^\circ$. 
The values of $\sipsi=15^\circ$ and $\silag=45^\circ$ are fixed across the pulse window.
Outside the loop the modelled PA track follows the proper mode at $\psi=90^\circ$,
because the assumed profile for the peak of the mixing angle distribution is: $\ppk=90^\circ -
60^\circ\exp(-0.5(\pi -\Phi)^2/\sigma^2)$. 
When $\ppk$ diverges from $90^\circ$ so much that the intermode is crossed
(dotted horizontal at $45^\circ$) the PA loop is opened on the
coproper OPM track.\footnote{It is therefore not necessary for the OPMs
left to the loop to be intermodal.}  
The loop is not identical to the observed one,
but several observed features are reproduced: there are upward-pointing `horns' of
PA at the top of the loop, little power inside the loop, and the bottom of
the loop extends into the downward-pointing tongue of radiative power that
reaches all the way to the top of the loop (after the PA axis is wrapped up with
the $180^\circ$ period). The twin minima in $L/I$ and the single-sign $V$
are also well reproduced. 

The simultaneous changes of $\lag$ seem to be indispensable in this model.
The change of $\psin$ alone does not produce the bifurcation, and the result
in such case resembles that of Fig.~11 in D17. 

The profile of peak value in the lag
distribution was $\lpk=360^\circ - 100^\circ\exp(-0.5(\pi -
\Phi)^2/\sigma^2)$, ie.~the amplitude of the lag change is equal to
$100^\circ$ in Fig.~\ref{pal}. 
When the lag-change amplitude is decreased
to $30^\circ$, the result of Fig.~\ref{palb} is obtained. One can see that
the loop disappears and is transformed into the U-shaped PA distortion, much
like the one observed in the data at $4.5$ GHz (Fig.~1 in MRA16). Moreover, the twin minima in
$L/I$ become shallower, but do not merge, again as observed for B1933$+$16.
The value of $L/I$ increases, while $|V|/I$ decreases, in agreement with
data at both $1.5$ GHz and $4.5$ GHz. 

The decrease of lag amplitude is thus the only thing which is needed to
understand the evolution of the PA loop  with frequency. It is possible to obtain the right behaviour with the
decrease of lag only, which is naturally expected at the increased $\nu$. 

The phase lag may be a strong function of $\nu$ and 
it is possible that $\lag$ becomes negligible at $4.5$ GHz. If so, then the
observed PA track of the U-shaped distortion directly reveals the profile of
$\ppk$,
which is indeed observed to be about $45^\circ$ away from the OPMs that are observed
outside the loop (the dotted curve in Fig.~\ref{palb}a is reproduced in the PA
curve of Fig.~\ref{palb}b, to be compared with the $4.5$ GHz data in Fig.~1 of MRA16).
 
The result of this section again shows that the core polarization of pulsars is a
combination of amplitude-driven and lag-driven effects, and the look of PA
curves and other polarization characteristics change with frequency, because
of the frequency-dependent phase lag. 
When the Gaussian profiles of $\ppk(\Phi)$ and $\lpk(\Phi)$ are misaligned,
the resulting profiles of $L/I$ and $V/I$ become asymmetric, which is
observed at both frequencies. It should be possible to construct a similar
multifrequency model for the complex behaviour of core polarization in J0437$-$4715
at different frequencies.

It must be noted that several effects of the lag change can also be produced
through the narrowing of $\nlag$ distribution. For example, a result similar
to that of Fig.~\ref{pal} can be obtained for a fixed ($\Phi$-independent)
$\lpk$ when $\silag$ is changing within the loop. Reasonably looking loops
were in particular obtained for a one-sided $\nlag$ with $\silag$ following the profile
of $\silag=145^\circ - 135^\circ\exp(-0.5(\pi-\Phi)^2/\sigma^2)$. 
In such case the exact shape of the resulting loop depends on the $\lpk$
value. 

\subsection{Lag-driven inversions of PA distortions}

\begin{figure}
\includegraphics[width=0.49\textwidth]{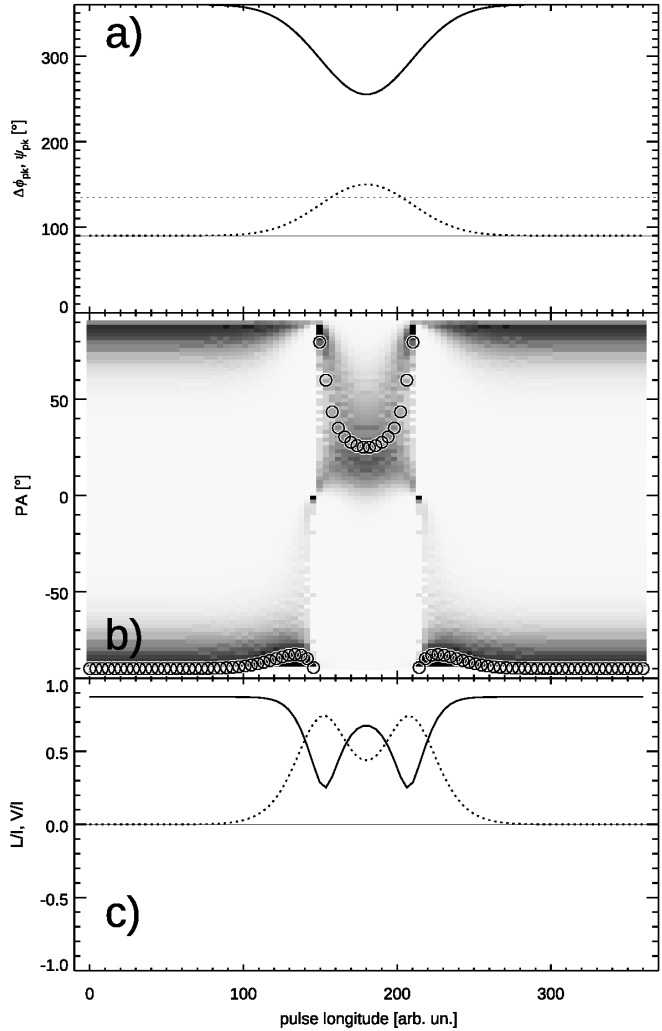}
\caption{
Polarization characteristics calculated for the $\ppk$ and $\lpk$ profiles
shown in panel {\bf a}, and for $\sipsi=15^\circ$, $\silag=45^\circ$. 
The PA is distorted downward, and $V$ is positive.
}
\label{prz}
\end{figure}

Polarization characteristics that result from coherent mode addition 
 sometimes are very sensitive to the parameters used. 
The results illustrated in the previous section were calculated for symmetric (two-sided) lag distribution $\nlag$. 
Fig.~\ref{prz} presents a different result for a one-sided $\nlag$. 
The amplitudes of the $\ppk$ and $\lpk$ profiles are $60^\circ$ and
$105^\circ$, respectively, and $\ppk$ now increases within the
PA distortion (see the dotted line in panel {\bf
a}). 

\begin{figure}
\includegraphics[width=0.49\textwidth]{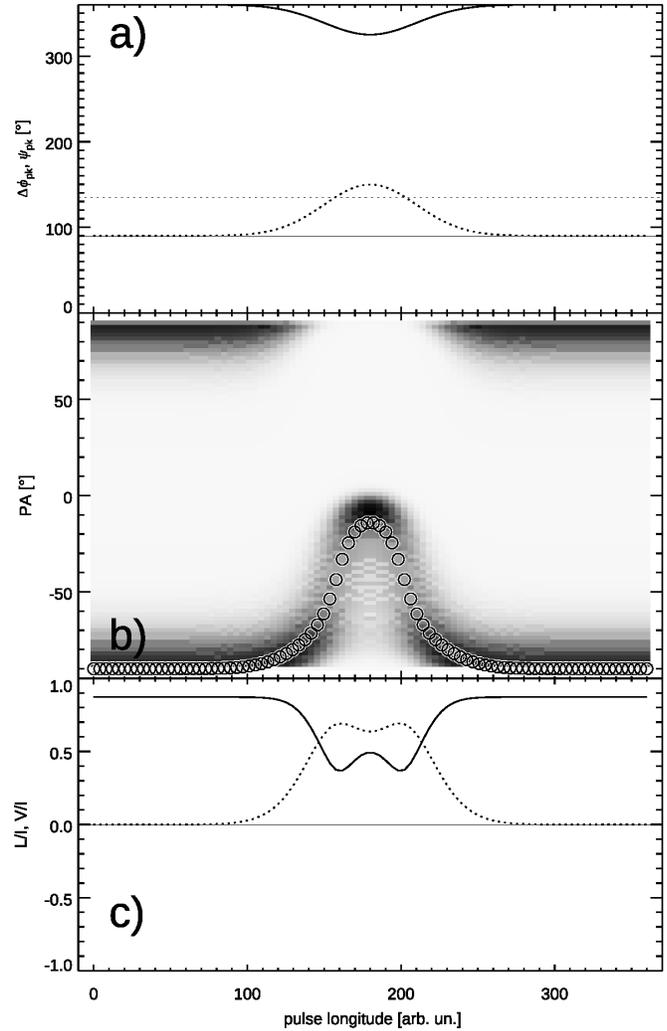}
\caption{Same as in previous figure, but for a smaller amplitude of 
the $\lpk$ profile (top solid line). 
The PA is now distorted upward, but the sign of $V$ does not change. This
behaviour resembles that observed in the core of B1237$+$25 in different
modulation states (Fig.~1 and 6 in SRM13).
}
\label{przb}
\end{figure}

A change of only the lag amplitude to $35^\circ$ leads to the result of
Fig.~\ref{przb}. The PA distortion is now protruding upwards, whereas the
other polarization characteristics (such as $L/I$ and $V/I$ do not change
much). This phenomenon resembles the PA bifurcation of B1237$+$25 in the N and Ab
modulation states (SRM13). With the change of modulation state, the observed PA follows
different branch of the PA bifurcation while the sign of $V$ does not change. 
It appears possible then, that the exchange of the followed PA branch is
caused only by the change of the lag value in different modulation states.


\begin{figure*}
\includegraphics[width=0.89\textwidth]{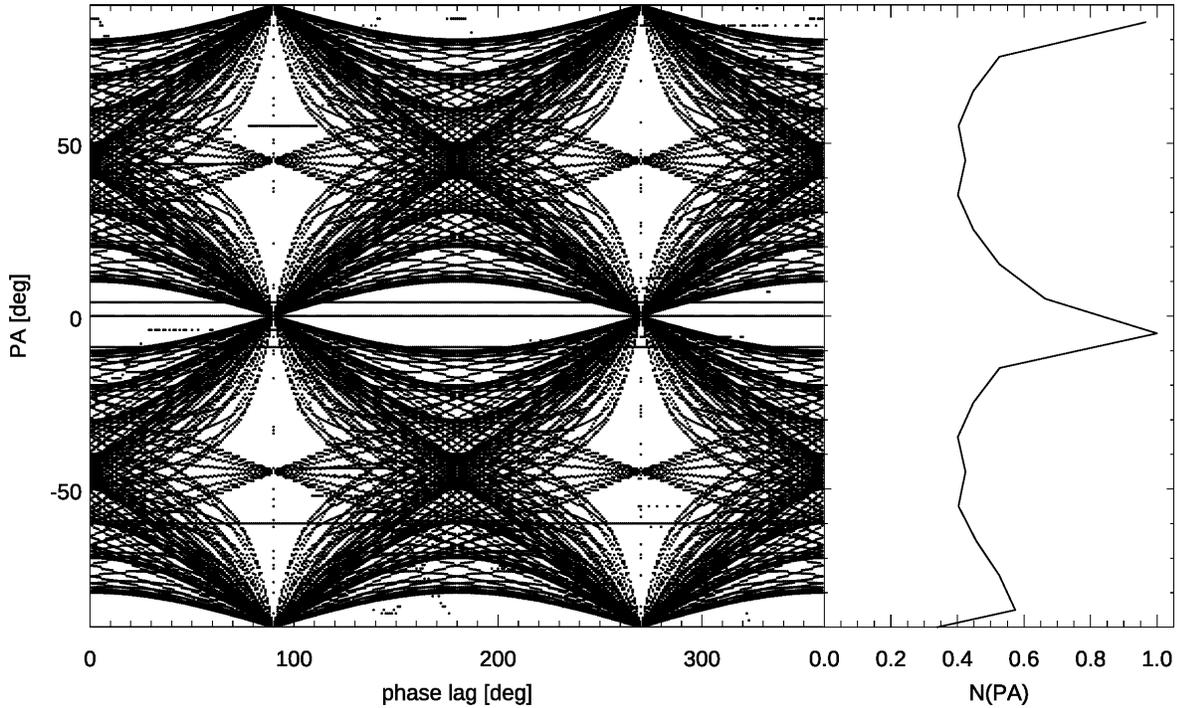}

\caption{The lag-PA diagram for sparsely and uniformly sampled parameter
space of  $(\lag,\mix,\beta)$. The result has been obtained for intervals:
$\Delta(\lag)=1^\circ$, $\Delta\mix=10^\circ$ and $\Delta\beta=5^\circ$. 
The coproper nodes at $\psi=n90^\circ$, the linearly polarized diagonals, and the
intermodes at $\psi=45^\circ+n90^\circ$ seem to stand out most in this picture.
For increasing $\lag$, a radio signal with fixed $\mix$ and $\beta$ would have
the PA changing along one of the visible lines. 
}
\label{mozaj}
\end{figure*}

Obviously, the phenomenon of the modulation-state-dependent polarization, and
other complex polarization phenomena in pulsars require more detailed study.
The parameter space for the coherent addition of non-equal elliptical modes offers a
large number of possible polarization profiles. Fig.~\ref{mozaj} 
presents the PA as a function of lag, calculated for a sparse grid of
parameters: $\Delta\mix=10^\circ$, $\Delta\beta=5^\circ$, and
$\Delta(\Delta\phi)=1^\circ$. Different lines of $\psi(\lag)$ correspond to
different pairs of $(\mix,\beta)$. A numerical code for pulsar
polarization needs to probe even larger parameter space, with the added widths of 
statistical distributions of $\mix$, $\lag$, and $\beta$. To make things more
complex, it may be necessary to introduce a few additional
parameters that describe how these six basic parameters depend on pulse longitude
$\Phi$.  More complete analysis of the phenomena visible in Fig.~\ref{mozaj}, along with
detailed numerical fitting of pulsar data, is deferred to further study.

\section{Conclusions}

It has been shown that complex non-RVM polarization effects in radio pulsars 
can be understood in geometrical terms, as the result of coherent and
quasi-coherent addition of elliptically polarized orthogonal proper-mode waves. 
The phenomenon of coherent mode addition is described by three (or six)
parameters: the phase lag, the amplitude ratio (mixing angle), and the
eccentricity of polarization ellipse (plus the widths of their
distributions). The model implies that the observed radio polarization is
driven by at least two independent effects: the changes of mode amplitude
ratio, which, in particular, are responsible for the regular OPM behaviour (with zero $V$ at
OPM transitions) and the changes of the phase lag which have opposite
characteristics. Both these factors influence the observed polarization
 within the same pulse intervals, which is evident in the core
region of profiles.
Such model explains several complicated and dissimilar 
phenomena, such as: distortions, bifurcations and loops of the PA observed in the central
part of profiles, twin minima in $L/I$
associated with these distortions, maxima of $|V|/I$ at OPM jumps, 
$45^\circ$-off PA tracks,
chaotic spread of PA values within the $45^\circ$-displaced emission, and dissimilar $L/I$ minima 
of mixed origin, such as those observed in J0437$-$4715. Moreover, the model is capable to 
interpret the changing look of these phenomena with frequency and possibly
with modulation state. 

The observed OPM tracks have often been directly associated with the natural
propagation mode waves. It has been shown here that the observed OPMs do
not necessarily correspond to the natural waves. Instead, the observed OPMs are a statistical average
of coherent sum of the natural waves (with diverse phase lags). Therefore,
the PA of observed polarization tracks can be completely different from the PA of the
natural waves. The observed PA tracks may be non-orthogonal and they
may wander away from the RVM PA. 
The coherent addition model implies that the PA is distorted by the
$\nu$-dependent location and width of the lag distribution, and by the 
$\nu$-dependent ratio of modal amplitude, as expressed by eq.~(\ref{psinon}).  

In the noncoherent model the observed PA can only jump by $90^\circ$ when
one mode becomes stronger than another. In the coherent mode addition model, 
the noncoherent condition is obtained by coherent summation of numerous natural mode waves
at diverse phase lags. This typically causes the coproper modes M$_1$ and
M$_2$ to stand out in the data. Preference of equal modal amplitudes,
however, makes the intermodes C$_1$ and C$_2$ most pronounced. 

Identical amplitudes of the natural propagation mode waves (m$_1$ and m$_2$) 
are automatically produced when the
waves are fed by a
circularly polarized signal.
The coherent addition model 
then implies that two pairs of observed OPM tracks may in general 
appear in pulsar profiles, and the pairs are separated by $45^\circ$. 
Just like the linear-fed coproper modes, the intermodal OPMs are pronouced when the phase lag distribution is
wide, which introduces many polarization ellipses that all share the same PA
of $45^\circ$ (or $-45^\circ$, see Fig.~\ref{cirprinc}). 

In the case of the linear-fed coproper modes, the psedomodal OPM jumps are 
produced when $\psin$ is passing through the intermodal value of $45^\circ$. 
$|V|/I$ is maximum at such OPM transitions. 
In the case of the circular-fed equal-amplitude OPMs, the regular OPM jumps
take place when the handedness of the feeding wave is changed. In the case
of the general model with the elliptical proper modes, the regular OPM jumps
occur in the usual way (when one mode becomes stronger than another, for
whatever reason). 

When the mode amplitude ratio slowly deviates from $1$, the observed PA
makes a
non-orthogonal passage between the intermodal PA value and the natural mode
PA, eg.~between $45^\circ$ and $90^\circ$. 
Such change of PA does not have to be precisely equal to $45^\circ$ 
given the possible simultaneous change of PA
caused by the RVM effect. Examples of such slow wandering of PA between the
OPM values can often be found in
pulsar data, eg.~on the trailing side of profile in J0437$-$4715,
Fig.~\ref{j0437}. 


The presented model solves several problems that appeared in the analysis of D17. 
The complex polarization in the core components of both normal and
millisecond pulsars can be understood as the result of simultaneous changes
of phase lag (with pulse longitude and frequency) and of the mode amount
ratio (which changes at least with pulse longitude). The change of lag with $\nu$ is
responsible for the different look of the PA loop in B1933$+$16 at $1.5$ and
$4.5$ GHz. If the profiles of lag and mode ratio are misaligned in pulse
longitude, it is possible to produce the dissimilar twin minima in $L/I$
as observed in J0437$-$4715.
The original two-parameter lag-PA diagram of D17 seemed to
clearly indicate where the observed OPMs are located, but it is found here that the observed
`modes' (PA tracks) 
in general do not coincide with the natural modes. They can be at any
distance from the RVM PA, they can be non-orthogonal, and they can be intermodal 
wherever the amplitude ratio is close to unity. 


The result of Fig.~\ref{pal} shows that a fairly simple underlying model
(see the Gaussian profiles in top panel) can produce the very complex effect
of the PA loop (panel {\bf b}). The coherent mode addition thus presents a
capable interpretive tool. However, the model contains many parameters: 
at least the lag, mixing angle, widths of their distributions, plus six 
parameters for their pulse-longitude dependence (amplitude, peak longitude, and the
width, in the case of a Gaussian). Even with the ellipticity ignored, this
makes up for ten free parameters. Moreover, some pairs of
the parameters (such as the lag and mixing angle, or the peak lag value and the
width of the lag distribution) are degenerate at least to some degree. 
Therefore, it is not easy to find the best fit parameters through a
hand-made sampling of the parameter space. Neither it is easy to break the
degeneracy. A possible way out is to consider the $\nu$ dependence of
modelled phenomena, which has helped us to break the $\psin$-$\lag$
degeneracy 
in the case of the loop in B1933$+$16. Modelling of the single pulse data
(distributions of PA, $L/I$ and $V/I$ at a fixed $\Phi$) may also prove
useful.  A need for a carefully designed fitting code is
apparent.

\section*{acknowledgements}
I thank Richard Manchester for the average pulse data on J0437$-$4715
(Parkes Observatory).
Plotting of polarized fractions for B1919$+$21 was possible thanks to the
public Arecibo Observatory data base provided by Dipanjan Mitra, Mihir Arjunwadkar, and
Joanna Rankin (MAR15). 
I appreciate comments on the manuscript from Bronek Rudak, discussions with Adam
Frankowski and I thank Wolfgang Sieber for words of encouragement. 
This work was supported by the grant 2017/25/B/ST9/00385 of the National Science
Centre, Poland.
\bibliographystyle{mn2e}

\section{Appendix}

\begin{figure*}
\includegraphics[width=0.89\textwidth]{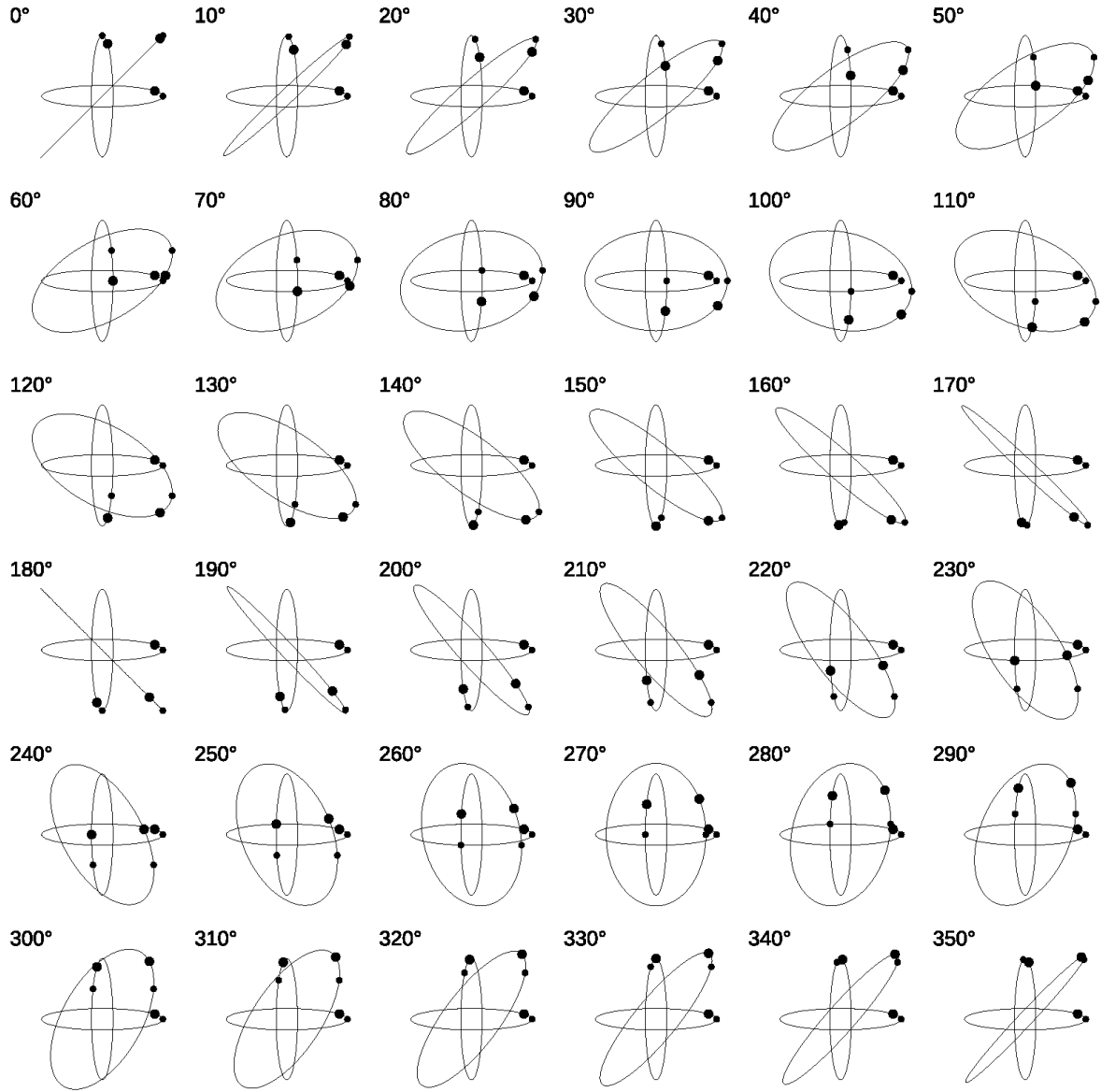}
\caption{Coherent summation of two orthogonal elliptically polarized natural
propagation mode waves (narrow orthogonal ellipses). Numbers in the corners
give the value of phase lag used in the summation.
The result has the form
of a wide ellipse,  a narrow ellipse at nearly diagonal orientation, or a
diagonal line (linear polarization at $\lag=0$ and $180^\circ$).
For most lag values the result has the PA close to $\pm45^\circ$. Only in the
second and fifth horizontal row the PA deviates considerably from
$\pm45^\circ$. The dots on the ellipses refer to the same moment of time
($t_1$ for small dots, $t_2 > t_1$ for the large dots, ie.~the direction of
electric field circulation is from a small dot to a large one). The
eccentricity angle of the ellipses that are added is $\beta=10^\circ$. The
result roughly corresponds to the S-shaped lines in Fig.~\ref{full3} and
\ref{full4}, ie.~it is not far from the discontinuous case with the PA jumping
from $45^\circ$ to $-45^\circ$ at $\lag=90^\circ$ (and back at $270^\circ$).
}
\label{full2a}
\end{figure*}

\begin{figure*}
\includegraphics[width=0.89\textwidth]{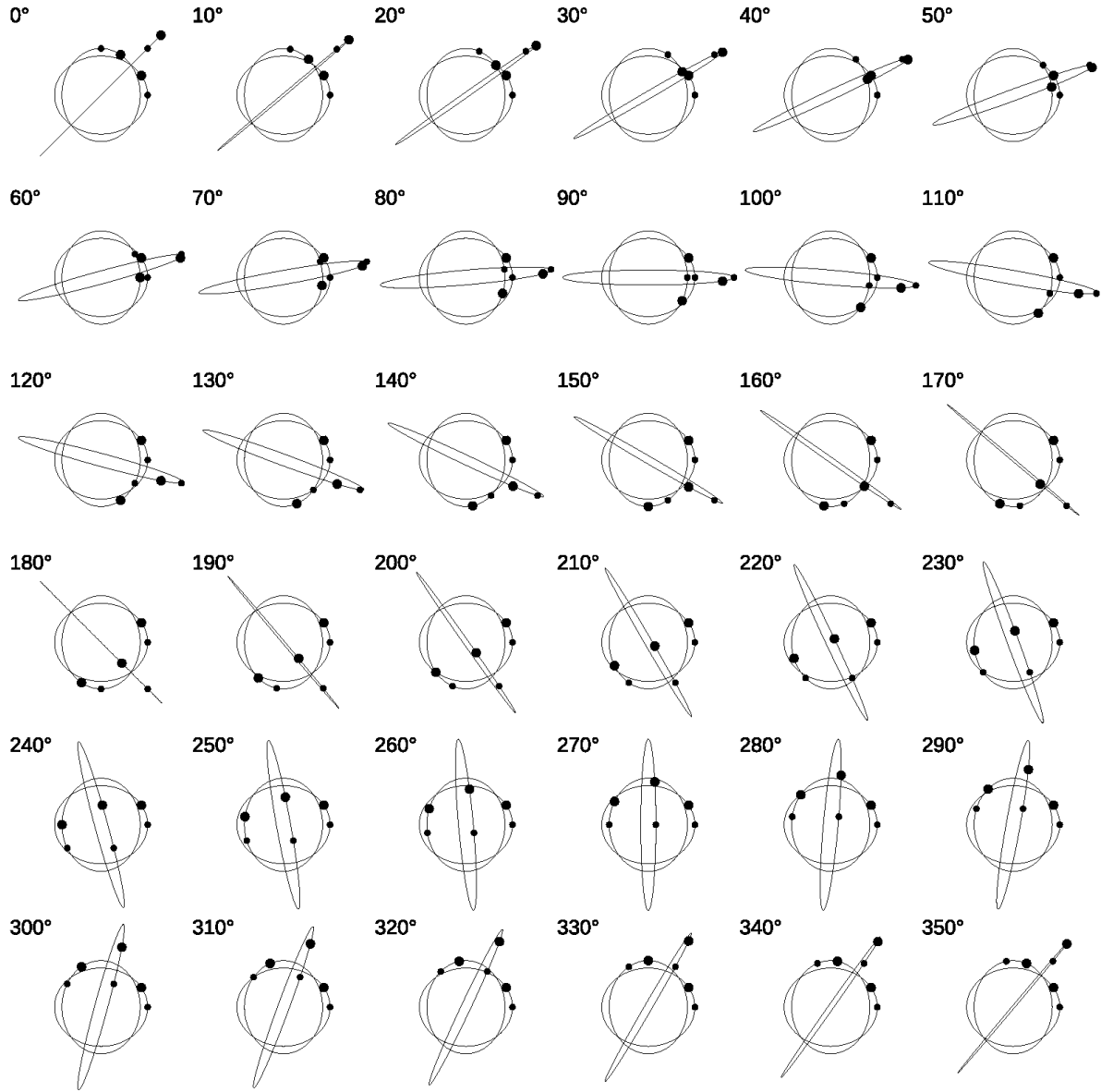}
\caption{
Same as in previous figure, but for nearly circularly polarized natural
waves ($\beta=40^\circ$). This case is close to the diagonals shown in
Figs.~\ref{full3} and \ref{full4}, ie.~the PA of the resulting wave increases
almost uniformly as in the Faraday rotation effect.
}
\label{full2b}
\end{figure*}



\bibliography{listofrefs2}


\end{document}